\documentclass[twocolumn,amssymb, amsmath, nobibnotes, nofootinbib, aps, superscriptaddress,pre]{revtex4-1}   

\usepackage{titlesec}
\usepackage{color, graphicx, hyperref, multirow, url, booktabs, colortbl, makecell, pbox, algorithm}

\usepackage[noend]{algpseudocode}

\algnewcommand{\To}{\textbf{To }}
\algnewcommand\Input{\item[\textbf{Input:}]}%
\algnewcommand\Output{\item[\textbf{Output:}]}%

\hypersetup{pdfnewwindow=true, colorlinks=true, linkcolor=red, citecolor=blue, filecolor=magenta, urlcolor=cyan}

\newcommand{\be}{\begin{enumerate}}
\newcommand{\ee}{\end{enumerate}}

\newcommand{\floor}[1]{\lfloor #1 \rfloor}

\newcommand{\overbar}[1]{\mkern 
  1.5mu\overline{\mkern-1.5mu#1\mkern-1.5mu}\mkern 1.5mu}
  
\newcommand\blfootnote[1]{%
	\begingroup
	\renewcommand\thefootnote{}\footnote{#1}%
	\addtocounter{footnote}{-1}%
	\endgroup
}

\usepackage{tikz}

\newcommand\PlaceText[3]{%
\begin{tikzpicture}[remember picture,overlay]
\node[outer sep=0pt,inner sep=0pt,anchor=south west] 
  at ([xshift=#1,yshift=-#2]current page.north west) {#3};
\end{tikzpicture}%
}

\begin{document}

\author{Upasana Dutta}
\email{upasanad@seas.upenn.edu}
\affiliation{Department of Computer Science, University of Colorado, Boulder, CO, USA}
\affiliation{Department of Computer and Information Science, University of Pennsylvania, Philadelphia, PA, USA}

\author{Bailey K. Fosdick}
\email{bailey.fosdick@cuanschutz.edu}
\affiliation{Department of Biostatistics and Informatics, University of Colorado Anschutz Medical Campus, Aurora, CO, USA}

\author{Aaron Clauset}
\email{aaron.clauset@colorado.edu}
\affiliation{Department of Computer Science, University of Colorado, Boulder, CO, USA}
\affiliation{BioFrontiers Institute, University of Colorado, Boulder, CO, USA}
\affiliation{Santa Fe Institute, Santa Fe, NM, USA}

\title{Sampling random graphs with specified degree sequences}

\begin{abstract}
The configuration model is a standard tool for uniformly generating random graphs with a specified degree sequence, and is often used as a null model to evaluate how much of an observed network's structure can be explained by its degree structure alone. A Markov chain Monte Carlo (MCMC) algorithm, based on a degree-preserving double-edge swap, provides an asymptotic solution to sample from the configuration model. However, accurately and efficiently detecting this Markov chain's convergence on its stationary distribution remains an unsolved problem. Here, we provide a  solution to detect convergence and sample from the configuration model. 
We develop an algorithm, based on the assortativity of the sampled graphs, for estimating the gap between effectively independent MCMC states, and a computationally efficient gap-estimation heuristic derived from analyzing a corpus of 509 empirical networks. We provide a convergence detection method based on the Dickey-Fuller Generalized Least Squares test, which we show is more accurate and efficient than three alternative Markov chain convergence tests.
\end{abstract}

\PlaceText{164.5mm}{37.3mm}{$^{,\hspace{0.5mm}\mathsection}$}

\maketitle

In the analysis and modeling of networks, random graph models are widely used as both a substrate for numerical experiments and as a null model or reference distribution to evaluate whether some network statistic is typical or unusual. \blfootnote{\hspace{-2.5mm}$^{\mathsection}$ These two authors contributed equally.} Given a sequence of non-negative integers $\{k\}$ whose sum is even, the configuration model aims at generating uniform random graphs with the degree sequence $\{k\}$.\ Thus, the configuration model is a special kind of random graph model, conditioned on a specified degree sequence, that allows researchers to assess the structural consequences of a network’s degree structure~\cite{molloy1998size, molloy1995critical, ring2020connected, newman2001random}. It is among the most widely used random graph models in network science~\cite{maslov2004detection, malmgren2010role, stouffer2007evidence, watts2007influentials, sundararajan2006local}, and it provides the basis for many theoretical results~\cite{chung2002average, miller2014epidemic, st2017susceptible, chatterjee2011random}.

\begin{figure}[b!]
\centering
    \includegraphics[width=0.48\textwidth]{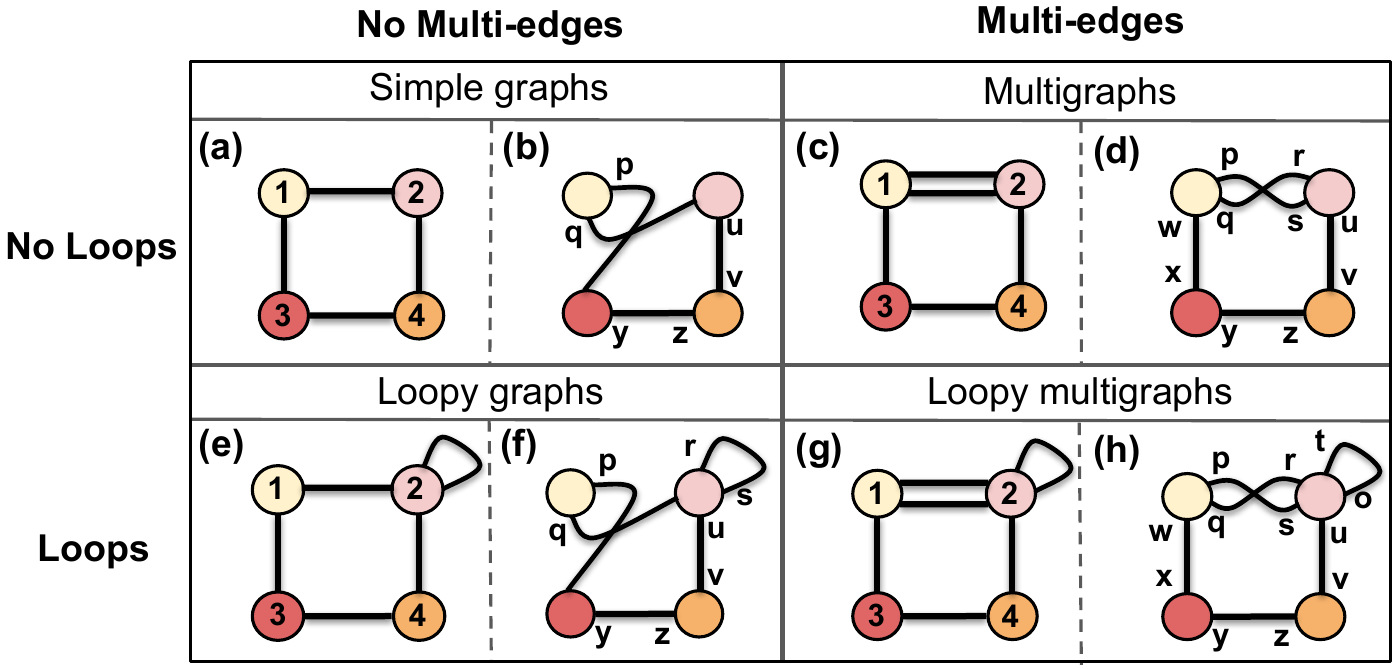}
  \caption{A small network in each of eight distinct configuration model graph spaces, corresponding to all combinations of allowing (c-d, g-h) or not allowing (a-b, e-f) multi-edges, and allowing (e-f, g-h) or not allowing (a-b, c-d) self-loops, in either a vertex-labeled space (1st and 3rd columns) or a stub-labeled space (2nd and 4th columns).}
    \label{EightGraphSpaces}
\end{figure}

The configuration model for networks with self-loops and multi-edges is the most well known~\cite{newman2018networks}. There are, in fact, eight different configuration models, depending on whether the random graph to be generated is vertex-labeled or stub-labeled---that is, whether or not it matters which “stub” on a vertex $i$ a “stub” from vertex $j$ attaches to---and whether it is allowed to have self-loops and/or multi-edges (Fig.~\ref{EightGraphSpaces}). The distinction between these different flavors of the configuration model has practical significance: Fosdick et al.~\cite{fosdick2018configuring} showed that the distribution of network statistics in different graph spaces can differ so much that an incorrect choice of graph space can lead to spurious or even opposite conclusions about the significance of some empirical networks' observed characteristics.

Sampling a graph from the configuration model is straightforward if the graph space is stub-labeled and allows the networks to have both self-loops and multi-edges. In this case, a simple stub matching algorithm suffices~\cite{newman2018networks}, which chooses a uniform random matching of all the edge stubs of the network, and can be run in $O(m)$ time, where $m$ is the number of edges. For all other graph spaces, it is common to instead generate a network using the stub matching algorithm, and then simply remove the self-loops or collapse the multi-edges in the generated network. As the graph grows, such self-loops and multi-edges are a vanishing fraction of all edges when the graph is sparse, as is usually desired. However, these modifications change the resulting network in highly non-random ways, because high-degree nodes are more likely to participate in both self-loops and multi-edges than are low-degree nodes. Hence, using the stub matching algorithm in this common way violates the underlying assumptions of the configuration model, and produces non-uniform draws from the target graph space. In other words: the stub-matching algorithm is theoretically justified for sampling only from the stub-labeled graph spaces where multi-edges and self-loops are allowed. 

An alternative solution for drawing from the other seven graph spaces, including simple graphs, is to sample them using a Markov chain Monte Carlo (MCMC) algorithm based on double-edge swaps. Although a number of such algorithms have been defined, the Fosdick et al. MCMC~\cite{fosdick2018configuring} is known to asymptotically converge on the target uniform distribution for each of the eight graph spaces (with rare exceptions in graph spaces that allow self-loops but not multi-edges~\cite{nishimura2018connectivity}). However, practical guidance on the time required for convergence with finite-sized networks remains unknown. This is the question we investigate and answer here. 

\textbf{Theoretical results for sampling from the configuration model:} The problem of sampling uniform random graphs with fixed degree sequence has been well studied. The earliest methods for generating uniform random networks with given degrees were based on the pairing model~\cite{bender1978asymptotic, bollobas1980probabilistic} (see~\cite{wormald1999models} for a brief history), which starts with a network where no stubs are connected, and then repeatedly chooses uniformly random pairs of stubs to connect forming an edge. If a simple graph is desired, the pairing algorithm is simply repeated until a simple graph is obtained. This method is impractical for generating simple networks because the probability of at lease one stub-match inducing a multi-edge or a self-loop is overwhelming as the size of the graph increases. For instance, this method would require about $10^{10}$ trials in expectation to produce one $k$-regular simple graph when $k = 10$ and the number of nodes in the network $n\rightarrow \infty$~\cite{bender1978asymptotic}. Algorithms for generating random graphs with specified properties, including degree sequence, have also been studied~\cite{tinhofer1979generation, tinhofer1990generating}, along with those based on the pairing model for sampling $k$-regular graphs in polynomial time when $k$ is bounded. These include sampling from an exactly uniform distribution when \hbox{$k$ = $O(\sqrt{\log n})$~\cite{wormald1984generating}}, and from asymptotically uniform distributions when \hbox{$k$ = $O(n^{1/3})$~\cite{sinclair1989approximate}}, $o(n^{1/5})$~\cite{frieze1988random}, $o(n^{1/11}(\log n)^{-3/11})$~\cite{steger1999generating}, and $o(n^{1/3})$~\cite{kim2003generating}.
In these algorithms, the output distribution becomes closer to uniform as $n$ gets larger, but there is no parameter for controlling how far the distribution is from uniform. 

A significant advancement on exactly uniform generation of simple graphs with more flexible degrees was made using a switch-based algorithm~\cite{mckay1990uniform}, which allowed sampling of exactly uniform random graphs in \hbox{$O(m + (\sum_ik_i^2)^2)$} average running time, when the maximum degree $k_{\max}$ satisfies the conditions $k_{\max}^3 = O(m^2/\sum_ik_i^2)$ and $k_{\max}^3 = o(m + \sum_ik_i^2)$. Here $m$ is the number of edges in the graph and $k_i$ is the degree of node $i$. For \hbox{$k$-regular} graphs, this algorithm runs in $O(n^2k^4)$ average running time when $k = O(n^{1/3})$. The algorithm generates random pairings from the pairing model resulting in both self-loops and multi-edges, and then uses random switchings to remove and add pairs of edges, reaching a simple graph in the end. Later, this algorithm was improved for generating exactly uniform $k$-regular simple graphs in running time $O(nk^3)$\cite{gao2017uniform} and $O(nk + k^4)$~\cite{arman2021fast} for the more relaxed bound of $k = o(\sqrt{n})$. This method was subsequently adapted to uniformly sample random graphs with power-law degree sequences with exponent $\gamma \geq 2.88$ in $O(n^{4.081})$~\cite{gao2018uniform} and $O(n)$\cite{arman2021fast} time.

More direct solutions for faster sampling of simple graphs with arbitrary degree sequence were proposed using sequential importance sampling with a $O(mk_{\text{max}})$ running time for $k_{\text{max}} = O(m^{1/4})$~\cite{bayati2010sequential}, thus improving the previous best known running time of $O(m^2k_{\text{max}}^2)$~\cite{mckay1990uniform}. Later, approaches for sampling general graphs were developed with a worst case running time of $O(n^2m)$~\cite{blitzstein2011sequential}. Similar methods for producing biased non-uniform samples, which are then re-weighted to compute unbiased estimates and probabilities of the desired uniform distribution have also been proposed, with a running time of $O(nm)$ for simple undirected~\cite{del2010efficient} and directed~\cite{kim2012constructing} graphs.\ However, there are no practical bounds on the number of samples required to achieve any particular desired accuracy against the target uniform distribution~\cite{fosdick2018configuring, zhao2013expand}, and deriving adequate bounds on the variance for importance sampling algorithms remains an open question~\cite{blitzstein2011sequential, bassetti2006examples}.

The first MCMC approach for sampling graphs with a given degree sequence was developed for approximate uniform sampling of any $k$-regular degree sequence~\cite{jerrum1990fast}. The Markov chain was proved to be rapidly mixing~\cite{aldous1983random, sinclair1988randomised}, taking time polynomial in $n$. However, because of the high order of the polynomial, the method had limited practical significance~\cite{wormald1999models}. This algorithm was extended to certain non-regular degree sequences~\cite{jerrum1990fast} for which the Markov chain's mixing time is polynomial only if the degree sequence is P-stable (see ~\cite{jerrum1989graphical, jerrum1990fast} for details on the conditions of P-stability). Intuitively, a degree sequence $\{k\}$ is P-stable if the number of possible graphs with the degree sequence $\{k\}$ does not change drastically when $\{k\}$ is slightly perturbed~\cite{jerrum1990fast}. Markov-chain based methods have also been proposed to sample from almost uniform distributions, which has been conjectured to be rapidly mixing for arbitrary degree sequence \hbox{(KTV conjecture)~\cite{kannan1999simple}}, but proved so only for the regular~\cite{kannan1999simple} and half-regular~\cite{erdos2010towards} bipartite case. Even though these Markov chains' mixing times are bounded by a polynomial in $n$, the exact asymptotic bounds are not known and it has not been proved that the chains mix rapidly in general.

Improved Monte Carlo algorithms based on importance sampling were proposed for sampling simple graphs with arbitrary degree sequences~\cite{snijders1991enumeration}, and are especially useful in the fields of social networks~\cite{wasserman1977random, moreno1938statistics, homans1950human} and ecology~\cite{connor1979assembly, struass1982statistical, wilson1987methods}. 
Switching-based MCMC algorithms were also proposed for generating random (0,1)-matrices with prescribed row and column sums~\cite{rao1996markov, roberts2000simple}. However, these methods either do not sample from exact uniform distributions or are computationally expensive even for small networks, e.g., $n=100$~\cite{snijders1991enumeration, rao1996markov, roberts2000simple}.\\

\vspace{-5mm}

More recent advancements in MCMC approaches offer theoretical upper bounds on mixing times for sampling graphs with specific types of degree distributions. For example, the mixing time of one Markov chain for sampling $k$-regular graphs is polynomially bounded as $O(k^{16}n^9\log(kn))$ for undirected~\cite{cooper2007sampling}, and $O(k^{26}n^{10}\log(kn))$ for directed graphs~\cite{greenhill2011polynomial}. These results were then extended to the non-regular case when $3 \leq k_{\max} \leq 1/4\sqrt{2m} $, with a mixing time of $O(k_{\max}^{14}(2m)^{10}\log(2m))$~\cite{greenhill2014switch}. In practice, this algorithm works for degree sequences that are not too far from regular. Even though these Markov chains could be made to sample from an exactly uniform distribution by running the chain sufficiently long, the proven bounds on their mixing times are too high for any practical use. Additionally, none of these results cover the case of generic degree sequences and hence do not apply universally.

\textbf{Practical approaches for assessing MCMC convergence:} Since the convergence rates of different MCMC algorithms on their target distributions differ considerably, analytically estimating the mixing times for arbitrary MCMCs is not possible~\cite{tierney1994markov, roberts1996geometric, brooks1998convergence}. Instead, convergence can be detected in an online fashion by examining the local statistics of the sequence of states the MCMC visits~\cite{roberts1998markov, cowles1996markov, brooks1998general}. Such convergence tests have a common structure: (1) a statistic calculated from each state the MCMC visits, (2) a choice of how often to sample from the MCMC, referred to here as the sampling gap $\eta$, and (3) a statistical test to assess when the calculated statistic has converged to its steady state, i.e., asymptotic distribution. Many such tests have been developed for general MCMCs~\cite{garren1993convergence, dixit2017mcmc, johnson1996studying, heidelberger1983simulation, liu1992variational, mykland1995regeneration, ritter1992facilitating, roberts1992convergence, yu1994rates, yu1994qlooking, zellner1995gibbs}, and several are included in popular Python and R~packages~\cite{gelman1992inference, raftery1995number, geweke1991evaluating}. However, these methods are not designed specifically for the configuration model and, as we will show, they do not perform well when applied to it. Detailed comparative analysis of several MCMC convergence detection techniques have highlighted their limitations with respect to their theoretical biases and practical implementations~\cite{brooks1998some, roy2020convergence, brooks1998convergence, el2006comparison, sinharay2003assessing}. Relatedly, Cowles and Carlin~\cite{cowles1996markov} study 13 different general convergence diagnostics and find that every method can fail to detect the type of convergence they were designed to identify.

Here, we develop an efficient and accurate convergence detection method specifically for the Fosdick et al. double-edge swap MCMC~\cite{fosdick2018configuring} for sampling from the configuration model. This method requires only an input degree sequence and a choice of graph space. First, we develop an algorithm for estimating a sampling gap between MCMC states so that the sampled states are effectively independent. We then apply this algorithm to a corpus of 509 real-world and semi-synthetic networks spanning all eight graph spaces, and distill the experimental results into a simple set of decision rules, based on empirical scaling laws,  for selecting the sampling gap automatically. We then specify a test for detecting MCMC convergence based on applying a Dickey-Fuller Generalized Least Squares (DFGLS) test to the degree assortativity values of the networks sampled by the MCMC, and show through a series of experiments that this test is effective at detecting convergence.

We then compare this method to several generic MCMC convergence detection methods and show that the DFGLS method is both more accurate and more efficient when applied to real-world networks. We also show that at the point the DFGLS method detects convergence of the sampled graphs' assortativity statistics, other widely-used network statistics, including clustering coefficient, average path length, and the number of triangles and squares, have also converged.

\section{Materials and Methods}
The Markov chain Monte Carlo algorithm described by Fosdick et al. guarantees that the resulting stationary distribution of the MCMC is uniform over graphs with the specified degree sequence and graph space (except for a rare subset of loopy graphs without multi-edges~\cite{nishimura2018connectivity}). A graph space is chosen by specifying (1) whether self-loops are allowed or not, (2) whether multi-edges are allowed or not, and (3) whether the graph is vertex- or stub-labeled (Fig.~\ref{EightGraphSpaces}). In a vertex-labeled graph, the vertices have distinct labels, while the stubs do not; whereas in a stub-labeled graph, each stub has a distinct label and hence each vertex can be identified by the unique set of stubs attached to it. For example, the stub-labeled graph with the exact same stub-connections as in Fig.~\ref{EightGraphSpaces}h, except where stub $p$ is attached to stub $r$, and $q$ to $s$, would be distinguishable from the stub-labeled graph in Fig.~\ref{EightGraphSpaces}h; but they would not be distinct in a vertex-labeled space as both would be the same graph shown in Fig.~\ref{EightGraphSpaces}g.
\begin{figure}[t!]
\centering
    \includegraphics[width=0.48\textwidth]{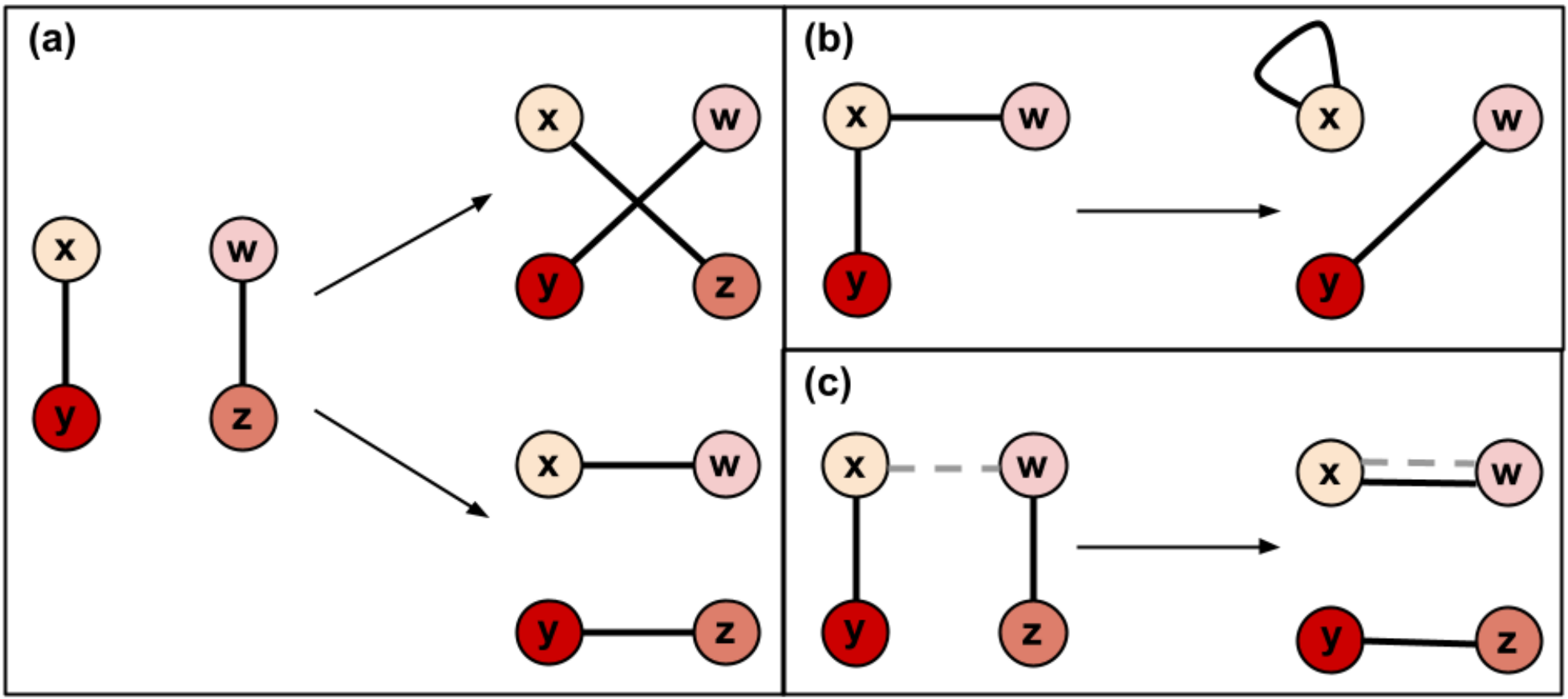}
  \caption{(a) The degree-preserving double-edge swaps on a pair of edges $\{(x, y), (w, z)\}$ results in either $\{(x, y), (w, z)\} \rightarrow \{(x, z), (w, y)\}$ or $\{(x, y), (w, z)\} \rightarrow \{(x, w), (y, z)\}$ as shown. (b) If the  vertices x, y, w, z are not distinct, double edge swap $\{(x, y), (w, x)\} \rightarrow \{(x, x), (w, y)\}$ can introduce a self-loop, (c) Similarly, if a third edge already exists among the vertices $x, y, w, z$, a double edge swap $\{(x, y), (w, z)\} \rightarrow \{(x, w), (y, z)\}$ can introduce a multi-edge.
}
\label{DoubleEdgeSwaps}
\end{figure}

A co-authorship network can be viewed as a vertex-labeled multigraph. If authors $A$ and $B$ have co-authored two papers, it would be nonsensical to match author $A$’s first collaboration stub with author $B$’s second collaboration stub or vice versa. In contrast, a network depicting an inter-school chess tournament is best viewed as stub-labeled with each node representing a school and stubs representing students in each school. Two students will be connected if they play a game against each other. Hence, every possible matching of stubs represents a distinct set of games among students, and such a network would be stub-labeled. See~\cite{fosdick2018configuring} for additional examples of vertex-labeled and stub-labeled networks.

In the MCMC we study here for sampling from the configuration model, each step of the chain is a degree-preserving double edge swap (Fig.~\ref{DoubleEdgeSwaps}a), which transitions from one graph $G_t$ to another graph \textrm{$G_{t+1}$}. At each step in the chain, we choose two edges $\{(x, y), (w, z)\}$ uniformly at random and replace them with either $\{(x, z), (w, y)\}$ or $\{(x, w), (y, z)\}$ with equal probability. Hence, a double-edge swap rewires exactly two edges, while preserving the degrees of all nodes in the graph.

In order to sample correctly from the target distribution, the space of graphs we choose imposes restrictions on which double-edge swaps are permitted. If the graph proposed by a particular swap is outside the specified graph space, e.g., it has a self-loop (Fig.~\ref{DoubleEdgeSwaps}b) or a multi-edge (Fig.~\ref{DoubleEdgeSwaps}c) when such connections are forbidden, or if certain other technical conditions are satisfied (see Appendix~\ref{appendix:B}), then the proposed change is rejected and the current graph $G_t$ is re-sampled, i.e., $G_{t+1} = G_t$. Because the Markov chain traverses a sequence of graphs where graph $G_t$ differs from $G_{t+1}$ by at most one double edge swap, graphs close to each other in the sequence are highly serially correlated. Additionally, if graphs are re-sampled very often, the serial correlation in the Markov chain is even higher. This correlation naturally decays between more distantly sampled points in the chain, but the distance required for the level of correlation to fall to a particular level must grow with the size of the network, because each double-edge swap changes at most a constant number of edges (two).


In some network studies, the sole purpose of using this MCMC is to estimate the mean value of some function of the network's structure, such as a network summary statistic like the clustering coefficient and average path length, or a statistic derived from more complicated network simulations, e.g., features of an SEIR network epidemiology simulation. In these cases, one analysis approach would be to compute the desired statistic on each graph visited by the Markov chain and continue sampling until a reasonable effective sample size (ESS)  is obtained.  Then, compute the mean summary statistic and its corresponding standard error using the ESS. This approach does not rely on drawing independent samples and therefore there is no need to estimate the gap necessary between two samples to make them effectively independent.\ However, this approach has two primary drawbacks.\ First, since only the \emph{a priori} desired statistic is stored in this procedure, if another statistic is later deemed of interest, the sampling algorithm would need to be re-executed, and the cost of this re-execution may be significantly greater, depending on the computational complexity of the new statistic of interest, e.g., degree assortativity is efficient to calculate in an online fashion, while most path-based statistics are not.

And second, the effective sample size grows extraordinarily slowly with the number of sampled graphs, because sequential steps in the Markov chain are highly correlated.\ As a result, if a researcher is storing every graph to make additional analyses subsequently, the memory requirement to store the graphs would be extremely prohibitive, even for networks of modest size (see \hbox{Appendix \ref{appendix:C})}.

The Fosdick et al. MCMC algorithm is guaranteed to sample graphs with a given degree sequence uniformly at random only after it has converged to its stationary distribution. Fosdick et al.\ provide a proof that this convergence is achieved in the asymptotic limit of $t \to \infty$. However, in practice, there exists some finite time $t_*$ at which the MCMC has effectively reached this asymptotic state. Our goal is to detect the earliest time at which this is true.

Here, we develop a complete solution to the practical task of sampling from the configuration model. Our solution divides this problem into three parts.

\begin{enumerate}
    \item We select a network-level summary statistic that quantifies a sufficiently non-trivial aspect of a network’s structure, so that we may transform a sequence of graphs $G_t$, $G_{t+1}$, $G_{t+2}$, \dots \enspace from the MCMC into a standard scalar time series $x_{t}$, $x_{t+1}$, $x_{t+2}$,\dots.
    \item We develop an algorithm for choosing a “sampling gap” $\eta_0$ for the given network such that the values $x_{t}$ and $x_{t+\eta_0}$ in the MCMC are statistically independent.
    \item Using a test of stationarity on thinned samples $\{x_t\}$, we assess the convergence of the MCMC on its stationary distribution.
\end{enumerate}

Throughout our analysis, we make extensive use of a corpus of real-world and semi-synthetic networks drawn from social, biological, technological domains~\cite{ICON}. We use these networks both to evaluate the methods we describe, and to develop a set of computationally lightweight, emperically grounded heuristics for automatically parameterizing our solution. These networks include 103 simple graphs, 154 loopy graphs, 142 multigraphs and 110 loopy multigraphs, and range in size from $n$ = 16 to 30,269 nodes with a variety of edge densities and degree distributions. Real networks with self-loops but no multi-edges and networks with multi-edges but no self-loops are relatively rare compared to those with both or neither of them. To obtain a sufficient number of such networks for our numerical experiments, we add or delete self-loops from empirical networks in our corpus to obtain semi-synthetic networks (see Sections \ref{LoopyResults} and \ref{MultigraphResults}).

Finally, to make the methods described here more accessible to the community, we provide our own implementations in a Python package, which can be found \href{https://upasanadutta98.github.io/ConfigModel_MCMC/}{here}.

\section{Results}
\subsection{Choosing the network statistic}
To characterize the progression of the MCMC through a graph space, we select the degree assortativity $r$ of a network as the cognizant  network summary statistic \hbox{$x_t = f(G_t)$}. The degree assortativity quantifies the tendency of nodes with similar degrees to be connected, and ranges over the interval $[-1, 1]$. Mathematically, $r$ is calculated as the normalized covariance of the degrees across all the edges of the network, given by,
\begin{equation}
    r = \displaystyle\frac{\sum_{xy} (A_{xy} - k_xk_y/2m)k_xk_y}{\sum_{xy} (k_x\delta(x, y) - k_xk_y/2m)k_xk_y}\enspace ,
\label{eq:1}
\end{equation}
where $A_{xy}$ is the adjacency matrix entry for nodes $x$ and $y$, $k_x$ is the degree of node $x$, $m$ is the number of edges in the network, and $\delta(x, y) = 1$ if $x = y$ and 0 otherwise.

The degree assortativity takes the value $r = 1$ if the graph is composed of only cliques, because in that case, the degree of every node is the same as that of its neighbors. An exception to this is when all the cliques in the network are of the same order, resulting in a $k$-regular network, for which the degree assortativity is undefined. The degree assortativity takes the value $r = -1$ if the graph is composed only of equal-sized stars, i.e., trees with exactly one internal node and $\ell \geq 2$ leaves. In that case, the highest degree nodes of the network connect only to nodes with the lowest degree. While it is common to expect that $r = 0$ in a random graph, for many graph spaces this is not the case~\cite{fosdick2018configuring}.

There are, of course, many alternative network-level summary statistics that could be used instead of degree assortativity, including the clustering coefficient, mean geodesic path length, mean betweenness centrality, and many more. However, many such statistics are computationally expensive to calculate repeatedly, for each step of the MCMC, which would limit the scalability of a sampling algorithm. The degree assortativity admits a computationally efficient update equation such that it can be calculated quickly after every double-edge swap of the Markov chain, allowing an algorithm to sample longer chains and larger networks.

Suppose that a double-edge swap $\{(x, y), (w, z)\}\rightarrow\{(x, w), (y, z)\}$ is performed (Fig.~\ref{DoubleEdgeSwaps}a). Using the definition of degree assortativity, its change from this swap can be written as

\begin{equation}
    \Delta r = \displaystyle\frac{(k_xk_w + k_yk_z - k_xk_y - k_wk_z)\times 4m}{(\sum_xk_x\times \sum_xk_x^3) - (\sum_xk_x^2)^2}\enspace ,
\label{eq:2}
\end{equation}
(see Appendix \ref{appendix:deriv_delta_r} for derivation).

Given a fixed degree sequence, the denominator in Eq.~\eqref{eq:2} is a constant and hence can be calculated once at a cost of $O(m)$ when the MCMC is first initialized, and stored for reference later. The numerator only requires the degrees $k_x$, $k_y$, $k_w$, $k_z$ of the four vertices involved in the swap, and the number of edges $m$. Hence it takes only constant time $O(1)$ to update $r$, given the assortativity of the current graph $G_t$ and the degrees of the nodes chosen for the double-edge swap. Calculating the initial assortativity $r_0$ is more expensive, but is done only once with a cost that amortizes over the length of the chain. It can be calculated as,
\begin{equation}
    r_0 = \displaystyle\frac{S_1S_\ell - S_2^2}{S_1S_3 - S_2^2}\enspace ,
\label{eq:3}
\end{equation}
where $S_1 = \sum_xk_x, S_2 = \sum_xk_x^2, S_3 = \sum_xk_x^3$ and $S_\ell = \sum_{xy}A_{xy}k_xk_y = 2 \sum_{(x,y)\in E}k_xk_y$. The expression in Eq.~\eqref{eq:3} contains only $O(n+m)$ terms and hence is substantially more efficient than Eq.~\eqref{eq:1}, which contains $O(n^2)$ terms~\cite{newman2018networks}. Only in the case of dense networks, where $m = \Theta(n^2)$, are the two calculations equally inefficient.

A caveat of choosing degree assortativity as the network statistic for monitoring the progression of the MCMC over time is that it is undefined for $k$-regular networks. For these networks, an alternative network statistic must be used, such as any of the ones mentioned above, with their corresponding higher computational cost.

\begin{figure}[t!]
\centering
    \includegraphics[width=0.48\textwidth]{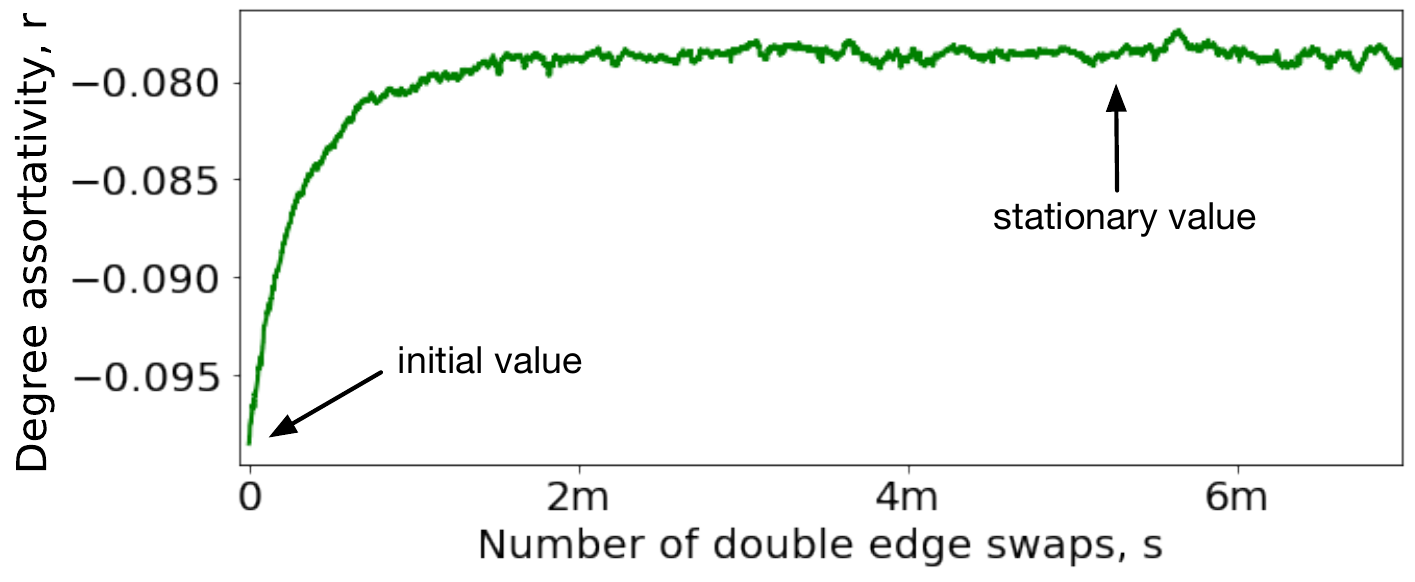}
  \caption{Network degree assortativity $r$ as a function of the number of double-edge swaps $s$ performed in the Markov chain, for a vertex-labeled simple graph with $n = 16,062$ nodes and $m = 25,593$ edges. The degree assortativity moves away from the initial value $r_0 = -0.098$ as the Markov chain progresses, and then converges towards a value, displaying non-trivial fluctuations around it.}
\label{ChangeIn_r}
\end{figure}

To illustrate the evolution of the degree assortativity $r$ over the course of a Markov chain, we apply the Fosdick et al. MCMC to a modest sized vertex-labeled simple graph with $n = 16,062$ nodes and $m = 25,593$ edges (Fig.~\ref{ChangeIn_r}). In the early part of the Markov chain, the degree assortativity quickly moves away from the initial value $r_0$ as the double-edge swaps initially randomize the empirical correlations in the network’s structure. In general, as a Markov chain progresses, the degree assortativity converges on some particular value, and then fluctuates around it. The key problem that convergence detection seeks to solve is deciding when statistical excursions are sufficiently random that we may declare the Markov chain to have reached its stationary distribution. In making this decision, we note that it is better to detect convergence too late rather than too early: a late decision merely wastes time in the form of extra steps in the Markov chain, while an early decision results in sampling from the wrong distribution of graphs.

\begin{algorithm}[H]
  \begin{algorithmic}[1]
    \Input{$G_0$ (network), $C$ (the number of independent MCMC chains), $S_T$ (list of degree assortativity values of size T), $\alpha$ (significance level for each test), $u$ (lower bound on number of MCMC chains that have significant lag-1 autocorrelation to reject independence) }
    \Output{The sampling gap $\eta_0$ for graph $G_0$}
      \State Let $m$ be the number of edges in $G_0$
      \State Run the MCMC for $1000m$ swaps (burn-in)
\State Let $x_1$ be the first degree assortativity value after the burn-in
\State $\eta = 0$
\State $d_\eta = C$
\While{$d_\eta > u$}  
\State $\eta = \eta + \floor{0.05m}$
\State $d_\eta = 0$
\For{$c \in [1, 2, \dots, C]$}                 
\State Construct $S_T$ s.t.\ for $1 \leq i \leq T$, $s_i\!= x_{\eta\times (i-1)+1}$

\State $d_c = $ CheckAutocorrLag1($S_T$, $\alpha$)
\State $d_\eta = d_\eta + d_c$
\EndFor

\EndWhile
\State \Return $\eta_0 = \eta$
\end{algorithmic}
\caption{\text{\tt GetSamplingGap}: Choosing the sampling gap $\eta_0$ for a network}
\label{Algorithm:1}
\end{algorithm}

\begin{algorithm}[H]
  \begin{algorithmic}[1]
    \Input{$S_T$ (time series of length T), $\alpha$ (significance level)}
    \Output{1, if lag-1 autocorrelation is statistically significant, and 0 otherwise}
    \State Let $\tau$ be the lag at which the sample autocorrelation is calculated
      \State $a =$ Autocorrelation($S_T$, $\tau = 1$)
      \State $\mu = -\displaystyle\frac{1}{T}$ \quad [Eq.~\eqref{eq:4}, with $\tau=1$]
      \State $\sigma^2 = \displaystyle\frac{T^4 - 4T^3 + 3T^2 + 4T - 4}{(T+1)T^2(T-1)^2}$ \quad [Eq.~\eqref{eq:5}, with $\tau=1$]
      \State $A = \displaystyle\frac{a - \mu}{\sigma}$
      \State z =  $(1-\alpha)$th quantile of N(0, 1).
      \If{$A > z$}
      \State \Return 1
      \Else 
      \State \Return 0
      \EndIf 
    \end{algorithmic}
\caption{\text{\tt CheckAutocorrLag1}: Test for significant lag-1 autocorrelation}
\label{Algorithm:2}
\end{algorithm}

\subsection{Choosing the sampling gap $\eta_0$}
By construction, the graphs that the Markov chain visits are serially correlated, as each double-edge swap changes at most four adjacencies. The magnitude of this serial correlation must therefore increase with the number of edges $m$, because it takes more steps in the Markov chain to randomize a large portion of the edges. As the task of detecting convergence is one of deciding when the fluctuations in the structure of the Markov chain's states are indicative of a stationary distribution, large serial correlations pose a significant problem by creating the appearance of non-random structure in the chain. To generate uniform random graphs from the configuration model, the sampled states from the MCMC must be sufficiently well separated so that they are effectively independent. To identify the spacing between states that yields a suitable sample, we develop an algorithm based on the autocorrelation function and statistical sampling theory, which can be applied to any network for obtaining an appropriate sampling gap $\eta_0$ between states. Given a choice of $\eta_0$, a sequence of sample states $X_{\eta_0} = (x_{t},x_{t+\eta_0},x_{t+2\eta_0},...,x_{t+(T-1)\eta_0})$ would then behave as a set of $T$ independent draws from the MCMC’s stationary distribution. 

The autocorrelation function of a time series measures the pairwise correlation of values $x_t$ and $x_{t+\tau}$ as a function of the lag $\tau$ that separates them (see Appendix \ref{appendix:A}). We can use a test for independence based on the autocorrelation function to determine whether a sample $X_{\eta}$, within which consecutive values are $\eta$ swaps apart in the Markov chain,  can be considered to be composed of independent and identically distributed (iid) draws from a stationary distribution.

The goal is to find the smallest value $\eta_0$ for which the sampled values in $X_{\eta}$ behave like iid draws from a stationary distribution. In a sample of $T$ iid normally distributed values, the autocorrelation value $a_\tau$ for any lag $\tau > 0$ has mean
\begin{equation}
    \mu(a_\tau) = -\displaystyle\frac{(T - \tau)}{T(T-1)}\enspace ,
\label{eq:4}
\end{equation}
and variance
\begin{equation}
    \sigma^2(a_\tau) = \displaystyle\frac{T^4 - (\tau + 3)T^3 + 3\tau T^2 + 2\tau(\tau+1)T - 4\tau^2}{(T+1)T^2(T-1)^2}\enspace ,
\label{eq:5}
\end{equation}
for $1 \leq \tau \leq T-1$~\cite{dufour1985some}. Hence, to assess if $X_{\eta}$ comprises uniform random draws from a stationary distribution (null hypothesis), we apply a hypothesis test that assesses the autocorrelation value of $X_{\eta}$ at $\tau = 1$, where the critical values are obtained from Eq.~\eqref{eq:4} and Eq.~\eqref{eq:5} (Algorithm \ref{Algorithm:2}) assuming a normal approximation as in Ref.~\cite{dufour1985some}. We use the above mean and variance equations for the sample autocorrelation instead of the more commonly used asymptotic normal approximation with mean zero and standard deviation $n^{-1/2}$ for all lags $\tau > 0$~\cite{box1970distribution, box2015time, brockwell2009time, brockwell2002introduction} because the asymptotic approximation has been shown to be a poor approximation of the true distribution of sample autocorrelation~\cite{kan2010distribution, dufour1985some}. We are particularly interested in the autocorrelation at lag $\tau = 1$ because each state in the MCMC is likely to be most correlated with the state immediately prior and immediately following it.\ For a sampling gap $\eta$, if the $\tau = 1$ autocorrelation of the sample $X_\eta$ is not statistically significant, the consecutive degree assortativity values in $X_\eta$ are effectively independent, and a sampling gap of $\eta$ is sufficient for drawing statistically uncorrelated states from the Markov chain.\ The hypothesis test in this case is one-sided (upper-tailed) since a state in the MCMC will always be positively correlated with the state exactly prior to it (except in pathological cases).

To initialize the algorithm (Algorithm \ref{Algorithm:1}), the MCMC is first run for $1000m$ swaps.\ This “burn-in" period ensures that in expectation, every edge has been proposed for a swap 2000 times, which we take as a reasonable degree of randomization before samples are taken for any experiment.\ After the burn-in period, for each choice of sampling gap $\eta$, the algorithm creates $C$ sequences from $C$ independent Markov chains.\ Each sequence is a list of the form $S_T = [s_1, s_2, s_3,\dots, s_T]$, where $s_i = x_{\eta(i-1)+1}$ for $1 \leq i \leq T$.\ We perform hypothesis tests on $C$ independent chains because a single Markov chain's trajectory may not be representative of the structure of the entire graph space. 

The number of chains $d_\eta$ for which a statistically significant $\tau=1$ autocorrelation is detected will tend to decrease with increasing sampling gap $\eta$. To account for multiple testing, we reject the null hypothesis of independence for the given sampling gap $\eta$ if more than $u$ tests are statistically significant, where $u$ is selected to control the family-wise error rate. The first value of $\eta$ at which $d_\eta \leq u$ is returned as the effective choice of sampling gap $\eta_0$ for the network. The upper bound $u$ is chosen based on the number of Markov chains $C$, the significance level $\alpha$ of each test, and the desired family-wise Type-I error. In our experiments, to obtain a family-wise Type-I error rate of about 5\% ($5.8\%$), we choose $C = 10$, $T = 500$, $\alpha = 0.04$, and $u = 1$. This allows our method to detect a $\tau = 1$ autocorrelation of 0.1 with 99.9\% power (see power-analysis in Appendix~\ref{appendix:A}).

Although the above sampling gap estimation algorithm can be applied to choose an appropriate value for $\eta_0$, the procedure itself is computationally expensive.\ To avoid this cost, we now develop a set of efficient heuristics and decision criteria using our corpus of empirical networks by which to automatically choose $\eta_0$ for a network, given its degree sequence and the choice of the graph space.

\subsubsection{\bf Simple graphs}

\begin{figure*}[t!]
\centering
    \includegraphics[width=\textwidth]{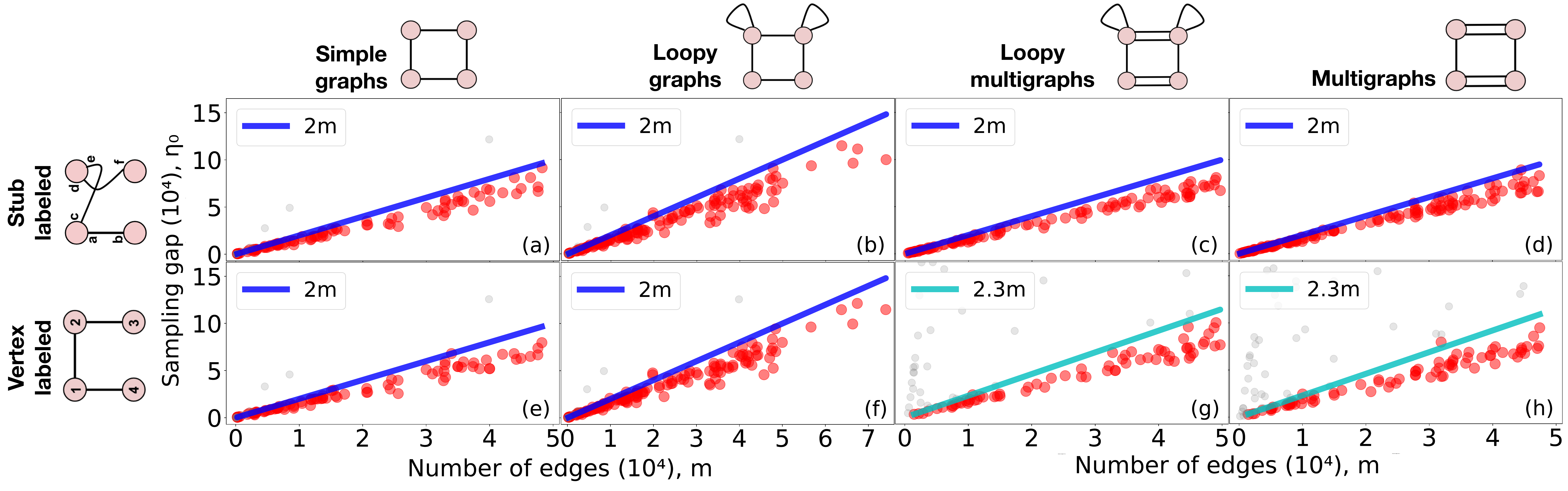}
  \caption{Space specific estimated sampling gap of networks for stub-labelled (a) simple graphs, (b) loopy graphs, (c) loopy multigraphs, (d) multigraphs, and vertex-labeled (e) simple graphs, (f) loopy graphs, (g) loopy multigraphs, and (h) multigraphs. The sampling gap $\eta_0$ of the networks that satisfy the density criterion ($\rho < 0.134$) in simple (a,e) and loopy (b,f) graph spaces, and the maximum degree criterion ($\max_i k_i^2 \leq 2m/3$) in the vertex-labeled loopy multigraph (g) and multigraph (h) spaces are shown as red circles, while those that do not satisfy these criteria, are shown in gray. Two networks in (a-b, e-f), 11 in (g) and 16 in (h) are not shown as their sampling gap fall above the plotted range.}
\label{SamplingGapResults}
\end{figure*}

For simple networks, the Markov chain’s transition probabilities are the same for both stub-labeled and vertex-labeled spaces (see Appendix \ref{appendix:simple}). Hence, the sampling gaps that Algorithm \ref{Algorithm:1} estimates for a simple network will be the same, in either the vertex-labeled and the stub-labeled spaces. 

We begin by running the sampling gap algorithm on each of the 103 simple networks in our corpus. The estimated sampling gap $\eta_0$ of the majority (86.4\%) of the networks satisfy an upper bound $\eta_0 = 2m$ (Fig.~\ref{SamplingGapResults}a,e). For the remaining networks (13.6\%), the estimated sampling gap $\eta_0$ (gray circles in Fig.~\ref{SamplingGapResults}a,e) is much larger than what this simple heuristic upper bound predicts. The reason the MCMC produces these unusually large $\eta_0$ estimates stems from the networks' high density and the Markov chain’s boundary enforcement criterion for simple graphs: if the Markov chain proposes a double-edge swap that would create a self-loop or a multi-edge, thereby exiting the specified graph space, the algorithm rejects this change and re-samples its current state. Repeated re-sampling extends the time needed for the serial correlation to decay. 

In order to estimate the likelihood of a re-sampling event in the Markov chain, we must consider the probability of the chain proposing an edge swap that would result in a multi-edge or self-loop. Suppose two edges $(x, y)$ and $(w, z)$ are chosen uniformly at random for a double-edge swap. We define two edges in a network to be adjacent to each other if they have exactly one common endpoint, implying that only 3 of the 4 vertices they are incident on are distinct. For a Markov chain to propose creating a multi-edge, the pair of chosen edges cannot be adjacent. The probability $q$ that two edges chosen uniformly at random are not adjacent is given by
\begin{equation}
    q = 1 - \displaystyle\frac{\displaystyle\Sigma_i{k_i^2} - 2m}{m^2 - m}\enspace .
\label{eq:q}
\end{equation}
\noindent Next, the two kinds of edge swaps (Fig.~\ref{DoubleEdgeSwaps}a) that can take place occur with equal probability. In both the cases, the probability of rejecting the swap due to the creation of a multi-edge is governed by the likelihood that an edge already exists where the swap proposes to place one. 

Thus, the probability of rejecting a proposed swap due to the creation of a multi-edge is given by
\begin{align}
&\textrm{Pr\big(rejection due to multi-edge\big)} \nonumber \\
&= q\times \Bigg[\displaystyle\frac{1}{2} \textrm{Pr\big(multi-edge}\big\rvert \{(x, y), (w, z)\}\rightarrow\{(x, z), (w, y)\}\big)\nonumber \\
&\hspace{0.45cm}+\displaystyle\frac{1}{2}\textrm{Pr\big(multi-edge}\big\rvert \{(x, y), (w, z)\}\rightarrow\{(x, w), (y, z)\}\big)\Bigg] \nonumber \\
&= q \times \Bigg[\displaystyle\frac{1}{2} \textrm{Pr\Big(at least one of }(x, z) \textrm{ and }(w, y)\textrm{ exists}\Big) \nonumber \\
&\hspace{0.45cm}+\displaystyle\frac{1}{2}\textrm{ Pr\Big(at least one of }(x, w)\textrm{ and }(y, z)\textrm{ exists}\Big)\Bigg] \nonumber \\
&\approx \displaystyle\frac{q}{2}\Big(1 - (1 - \rho)^2\Big) + \displaystyle\frac{q}{2}\Big(1 - (1 - \rho)^2\Big) \nonumber \\
&= \hspace{0.1cm} q \times (2\rho - \rho^2) \nonumber \\
&\approx 2\rho - \rho^2\enspace ,
\label{eq:8}
\end{align}
\noindent where the probability of an edge existing between two randomly chosen nodes is approximated as the network's edge density $\rho$. For simple graphs $\rho = \langle k\rangle/n-1$ and for loopy graphs $\rho = \langle k\rangle/n$, where $\langle k\rangle$ is the mean degree.

For all 103 simple networks in our corpus we find $q \approx 1$ (see Appendix~\ref{appendix:Rejection_multi_vs_loopy} Fig.~\ref{Rejection_multi_vs_loopy}a), meaning that the probability that two randomly chosen edges are non-adjacent to each other is approximately 1. This fact implies that the probability of rejecting a double-edge swap because it would introduce a multi-edge is approximately $2\rho -  \rho^2$. We refer to the simplified expression of $2\rho -  \rho^2  = \omega$ as the ``density factor" of a network, which gives a simple density-based estimate of Pr(rejection due to multi-edge) in a graph space that does not allow multi-edges. A more precise estimate would exploit the moment structure of the degree distribution, but such a formula is not necessary for our purposes. 

It can also be shown that the probability of rejecting a swap because it would introduce a self-loop (see Appendix~\ref{appendix:Rejection_multi_vs_loopy}) is $(1 - q)/2 \approx 0$, because $q\approx 1$. This fact implies that, the sampling behaviour of the Markov chain is governed more strongly by the rejection rate due to forming multi-edges than from forming self-loops (see Appendix~\ref{appendix:Rejection_multi_vs_loopy} Fig.~\ref{Rejection_multi_vs_loopy}b), and this rate is governed by the network's density. 

In our corpus of 103 simple networks, the vast majority (86.4\%) have a density factor $\omega < 0.25$, six (5.8\%) have $0.25 \leq \omega \leq 0.5$, five (4.9\%) have $0.5 < \omega \leq 0.75$ and the remaining four (3.9\%) have $\omega > 0.75$. These frequencies reflect the fact that real-world networks with very high density typically occur only rarely~\cite{broido2019scale}, unless the network is especially small. Across our simple network corpus, we find that networks with a density factor $\omega < 0.25$ yield sampling gap estimates that scale linearly with the number of edges $m$, while networks with a higher density are more likely to yield anomalously high sampling gaps. We use this observation to divide networks into two categories, based on their calculated densities $\rho$. If $\omega \geq 0.25$ for a network, implying that a network’s density $\rho \geq 0.134$, the sampling gap should be estimated via the algorithm described above (Algorithm \ref{Algorithm:1}); otherwise, a reasonable sampling gap is simply $\eta_0 = 2m$. Note that this provides an upper bound on the estimated gaps of nearly all the empirical networks (Fig.~\ref{SamplingGapResults} a,e).
There are four networks (4.5\%) that satisfy the network density criterion ($\rho < 0.134$) and yet their estimated sampling gap is on average 1.09 and 1.12 times our recommended upper bound $\eta_0 = 2m$ in the stub- and vertex-labeled spaces, respectively. 

\subsubsection{\bf Loopy graphs}
\label{LoopyResults}

Real-world ``loopy” networks, meaning networks with self-loops but no multi-edges, are uncommon. We construct a reasonable empirical corpus of 154 loopy graphs by obtaining 51 such networks from public repositories~\cite{ICON} and generating an additional 103 loopy graphs by adding self-loops to our previous corpus of simple networks. To convert a simple network into a loopy network, we first measured the fraction of nodes with a self-loop in both our corpus of 110 real-world loopy multigraphs and the 51 real-world loopy graphs to obtain an empirical distribution of loopiness.  We note that this loopiness distribution is strongly bimodal: 44\% have at most 5\% of nodes with self-loops, while 39.1\% have at least 95\% of nodes with self-loops. For each simple graph in our corpus, we then chose a random fraction from this empirical loopiness distribution and added a self-loop to each node with the corresponding probability.

Although the MCMC algorithm proposed by Fosdick et al. correctly samples graphs with fixed degree sequence post convergence, there are certain technical conditions~\cite{nishimura2018connectivity} that the degree sequence must satisfy for this to hold true in a loopy graph space in order for the double-edge swap algorithm to be able to reach every valid loopy graph with the given degree sequence. Only a rare set of loopy graphs fail to satisfy the required conditions, and in our corpus of loopy graphs, all the 154 networks' degree sequences met them. If a network fails to meet the required conditions, the Nishimura MCMC must instead be used~\cite{nishimura2018connectivity}, which augments double-edge swaps with triangle-loop swaps, rather than the Fosdick et al. MCMC used here.

For our loopy network corpus, we repeat our analysis by first running the sampling gap algorithm on each network in our loopy corpus. We note that because multi-edges are not permitted in the loopy space, the density criterion $\rho < 0.134$ has the same relevance here as it does with simple graphs. In our loopy corpus of 154 networks, 140 (90.9\%) networks have a density $\rho < 0.134$, and the same scaling law of $\eta_0 = 2m$ is an upper bound in both stub- and vertex-labeled spaces (see Appendix~\ref{appendix:loopy}) on nearly all estimated sampling gap values for these networks (Fig.~\ref{SamplingGapResults}b,f). Of the 140 loopy networks that have $\rho < 0.134$, there are three (2.1\%) and six (4.2\%) in the stub- and vertex-labeled spaces for which the estimated sampling gaps are on average 1.06 and 1.08 times our proposed upper bound $\eta_0 = 2m$, respectively.

\subsubsection{\bf Loopy multigraphs}

For stub-labeled loopy multigraphs, it is common to directly construct networks with a fixed degree sequence by choosing a uniformly random matching on the set of edge ``stubs” given by that sequence~\cite{newman2018networks,blitzstein2011sequential}. This algorithm is computationally cheap, running in $O(m)$ time, compared to an MCMC approach. However, stub matching cannot be used to correctly sample from the vertex-labeled loopy multigraph space. In that case, we instead use the Fosdick et al. MCMC. For completeness, we analyze the MCMC’s behavior in both the stub-labeled and vertex-labeled loopy multigraph settings.

In the loopy multigraph space, the Markov chain’s transition probabilities in the stub-labeled and vertex-labeled spaces are different. In the stub-labeled case, the Markov chain never re-samples a state, while in the vertex-labeled space, some of the proposed transitions are accepted with a probability less than 1. As a result, the Markov chain may re-sample some states often (see Appendix \ref{appendix:loopymultigraphs}), thus making the estimated sampling gap in the vertex-labeled space typically greater than in the stub-labeled space.

We investigate differences between the spaces by running the sampling gap algorithm on our corpus of 110 real-world loppy multigraphs (Fig.~\ref{SamplingGapResults}c,g). We find that the estimated sampling gap tends to be marginally higher for vertex-labeled graphs. However, for 42 of 110 vertex-labeled loopy multigraphs, the estimated sampling gap (gray circles in Fig.~\ref{SamplingGapResults}g) is much higher than the simple heuristic $\eta_0 = 2.3m$ that fits the majority of empirical networks in this graph space. This behaviour for these networks can be understood by considering how the edge multiplicities of a network and its degree distribution influence the MCMC’s dynamics.

In the vertex-labeled space, the likelihood of a swap being rejected is directly proportional to the multiplicity of the edges chosen for the swap (Appendix \ref{appendix:loopymultigraphs}). If some nodes of the network have very high degrees, the networks in the Markov chain can develop high edge-multiplicities over time, which increases the probability that a swap is rejected, leading to a $\Delta r = 0$ swap. A second issue arises when two edges of the network $(x, y)$ and $(w, z)$ share exactly one common endpoint. If a double edge swap proposes the change $\{(x, y), (w, z)\} \rightarrow \{(x, w), (y, z)\}$, then in expectation half the time the degree assortativity will remain unchanged after the swap, i.e., $\Delta r = 0$. Because these edges are chosen uniformly at random, the higher the degree of a node, the greater the probability of that node appearing twice in the group $(x, y, w, z)$. Regardless of the source, the more $\Delta r = 0$ steps that occur, the greater the serial correlation in the chain, and the greater the sampling gap required to obtain independent samples from the Markov chain.

To construct a heuristic that can efficiently decide when a graph is likely to produce such undesirable effects in the Markov chain, we draw on a related insight from Chung-Lu random graphs, which are a vertex-labeled model of simple graphs that constrain a network's maximum degree. For Chung-Lu graphs, this constraint limits the likelihood of producing a network with multi-edges, and has a form $\max_i k_i^2 \leq 2m$~\cite{chung2002connected}. In our setting, it provides the basis for a simple heuristic to decide whether a degree distribution is sufficiently right-skewed that it would generate overly correlated states in the Markov chain in the vertex-labeled space.  We find that every loopy multigraph in our corpus that exceeds a maximum degree criterion of $\max_i k_i^2 \leq 2m/3$ does indeed require an unusually large sampling gap in the vertex-labeled space (gray circles in Fig.~\ref{SamplingGapResults}g). In contrast, networks that fall below this threshold produce sampling gaps that exhibit the same linear relationship with $m$ observed in all graph spaces, and a scaling law of  $\eta_0 = 2.3m$ produces a conservative upper bound (Fig.~\ref{SamplingGapResults}g). In general, if a network exceeds a maximum degree criterion of $\max_i k_i^2 \leq 2m/3$, the sampling gap $\eta_0$ in the vertex-labeled space should be estimated using the sampling gap algorithm (Algorithm~\ref{Algorithm:1}). 

In the stub-labeled case, the corresponding scaling law obtained from all the 110 loopy multigraphs is \mbox{$\eta_0 = 2m$} (Fig.~\ref{SamplingGapResults}c). None of the networks in the stub-labeled space have $\eta_0$ higher than this proposed upper bound.

\begin{figure*}[t!]
\centering
    \includegraphics[width=\textwidth]{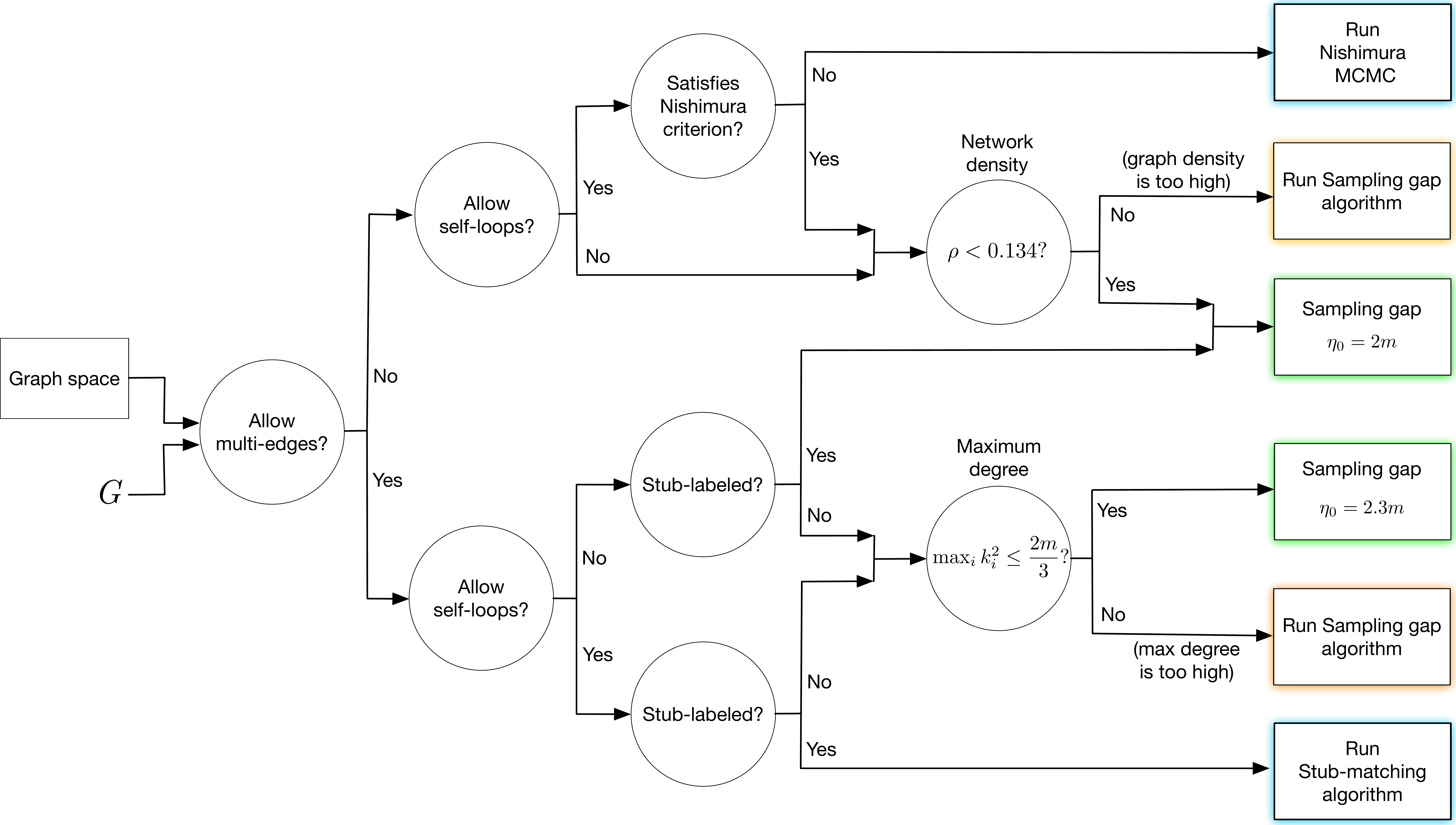}
  \caption{Decision tree of the space specific sampling gaps given by our heuristics, depending on the satisfiability of various constraints (see text). In the loopy graph space, when the Nishimura criterion~\cite{nishimura2018connectivity} is not satisfied, the Nishimura MCMC should be used instead of the Fosdick et al. MCMC. Similarly, for stub-labeled loopy multigraphs the stub-matching algorithm should be used.}
\label{DecisionTree}
\end{figure*}

\subsubsection{\bf Multigraphs}
\label{MultigraphResults}
Real world networks with multi-edges but no self-loops are also uncommon. We construct an empirical corpus of 142 such networks by obtaining 32 of them from public repositories and generating 110 multigraph networks by removing the self-loops of our loopy multigraphs. As with loopy multigraphs, the network’s degree distribution and whether the space is stub-labeled or vertex-labeled govern the choice of sampling gap $\eta_0$ (see Appendix~\ref{appendix:multigraphs}). Repeating our analysis on this corpus of 142 multigraphs, we find that the estimated sampling gaps of the networks in the stub- and vertex-labeled spaces (Fig.~\ref{SamplingGapResults}d,h) exhibit similar patterns to the loopy multigraph spaces. These results corroborate the fact that the probability of a double-edge swap rejection due to self-loops (Eq.~\eqref{eq:q}) in negligible, as reflected by the similarity of the Markov chain's behaviour in the multigraph and the loopy multigraph spaces. The scaling laws of $\eta_0 = 2m$ for loopy multigraphs in the stub-labeled space, and of $\eta_0 = 2.3m$ for loopy multigraphs in the vertex-labeled space that satisfy the maximum degree criterion $\max_i k_i^2 \leq 2m/3$ provide good upper bound on the estimated sampling gaps. Only one network (0.7\%) in the stub-labeled space has an estimated sampling gap 1.02 times our upper bound of $\eta_0 = 2m$, and two in the vertex-labeled space that satisfy the maximum degree criterion (2.5\%) have an estimated sampling gap $1.07$ times our upper bound of $\eta_0 = 2.3m$.

\subsection{Choosing the sampling gap efficiently}

Although a suitable sampling gap $\eta_0$ can be estimated using Algorithm~\ref{Algorithm:1} for any graph space and degree sequence, our numerical experiments show that in many cases, a sufficient gap may be chosen more efficiently using a simple scaling law in the number of edges in the network (Fig.~\ref{SamplingGapResults}). There are some conditions on applying these scaling laws in practice, depending on the particular graph space and certain structural properties of the network. In some cases, it is still be necessary to choose $\eta_0$ via Algorithm~\ref{Algorithm:1}. 

The decision tree in Fig.~\ref{DecisionTree} organizes the insights, conditions, and scaling laws obtained from our numerical experiments into a simple and efficient heuristic for choosing the sampling gap $\eta_0$, depending on the network’s properties and specified graph space. In cases where a network satisfies the constraints for which a scaling law can be used, we can choose $\eta_0$ directly. For instance, for each of the 89 networks in our simple network corpus that satisfy the density criterion ($\rho < 0.134$), we can immediately choose their sampling gap as $\eta_0 = 2m$, rather than running Algorithm~\ref{Algorithm:1}, saving substantial substantial computational time, ranging from a few seconds $(n=64, m=243)$ to about 30 minutes $(n=16,840, m=48,232)$ for the largest network in the corpus.

\subsection{Convergence detection}

In probability theory, the mixing time of a Markov chain is the number of steps the chain needs to run before its distance from stationarity is small~\cite{levin2017markov}. For practical purposes, the mixing time of a Markov chain often determines the run-time of the process that uses the Markov chain for sampling purposes. Detecting convergence is the practical task of deciding when a Markov chain has run sufficiently long to be well mixed.

\subsubsection{\bf Designing the convergence method}

We now specify a convergence test for the Fosdick et al. MCMC. This test uses the Dickey Fuller-Generalised Least Squares (DFGLS) test \cite{elliott1992efficient} to assess whether a sequence of states from the MCMC, represented as a sequence of degree assortativity values, possesses a unit root, against the alternative of stationarity. A time-series is called stationary when the statistical properties of the series, such as the mean, variance and the covariance, are independent of time. The DFGLS test first transforms the time-series via a generalised least square regression and then performs the Augmented Dickey Fuller (ADF) test \cite{dickey1979distribution} to test for stationarity. The DFGLS test has been shown to have greater statistical power than the ADF test \cite{elliott1992efficient}. We use the presence of stationarity in the Markov chain as evidence that the Markov chain has converged on its equilibrium.

To perform the DFGLS test of convergence, we first populate a list of degree assortativity values for a sequence of graphs sampled by the Markov chain, starting with the original network. 
\begin{figure}[t!]
\centering
    \includegraphics[width=0.48\textwidth]{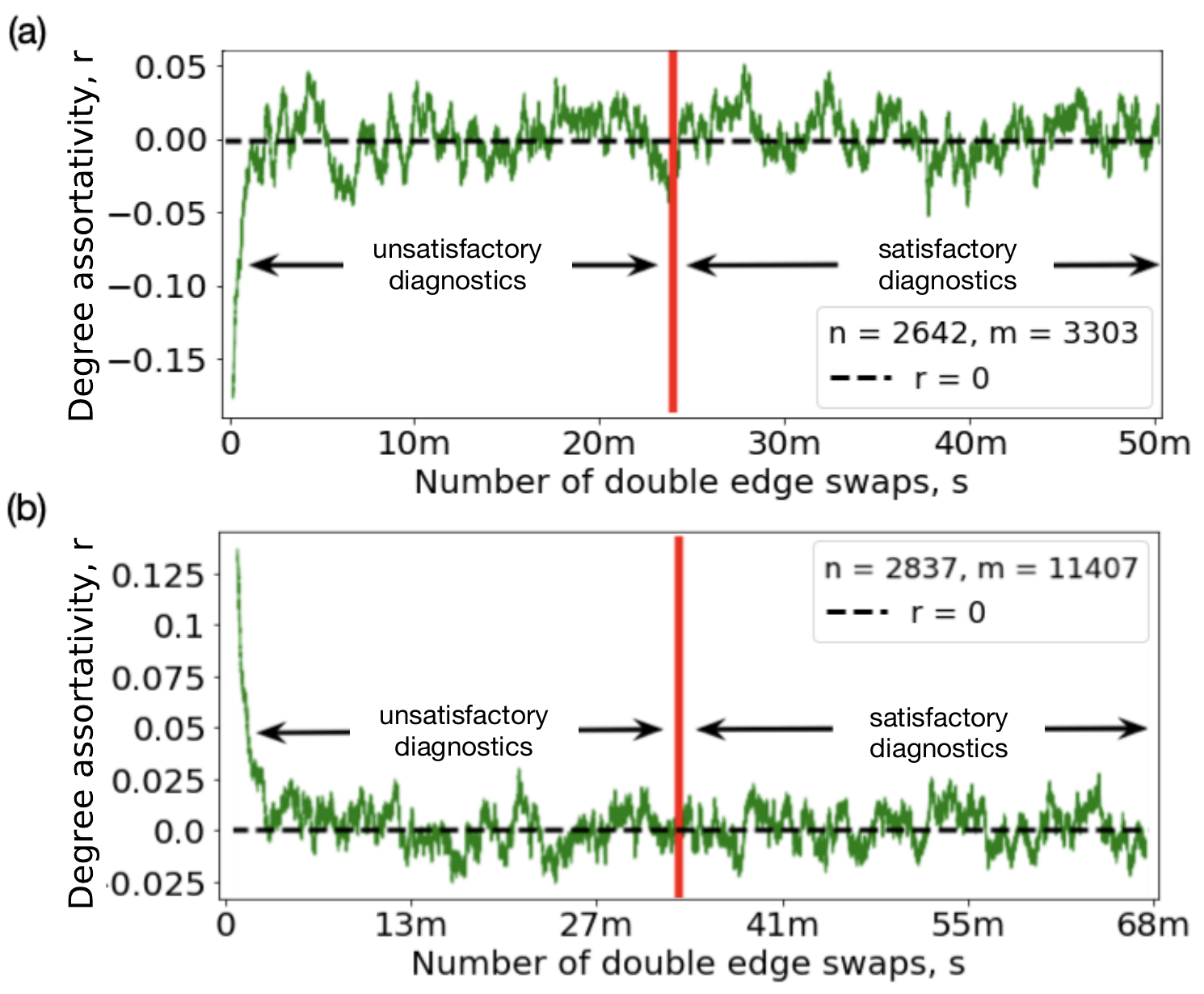}
  \caption{Network degree assortativity $r$ as a function of the number of double-edge swaps, $s$ performed in the Markov chain, for (a) a stub-labeled simple graph with $n = 2\,642$ nodes and $m = 3\,303$ edges, and (b) a vertex-labeled loopy multigraph graph with $n = 2\,837$ nodes and $m = 11,407$ edges. The vertical red line indicates the point at which our diagnostic suggests convergence.}
\label{ConvergenceDetection_green}
\end{figure}
We then perform the DFGLS test on the sampled list. The test determines whether the properties of the process generating the degree assortativities is changing over time. If the test does not reject the null-hypothesis of non-stationarity, we discard the contents of the list and ``slide" it forward, sample new degree assortativity values from the Markov chain to populate the list afresh, and then test again. We repeat this process of slide, repopulate, and test, until the DFGLS test rejects the null hypothesis of non-stationarity, which we interpret as an indicator that the MCMC has converged to its stationary distribution. We choose the length of this list, which we denote as the window-size, to be the sampling gap of the network. We choose the window-size to equal the sampling gap of the network because the sampling gap is proportional to the rate at which the MCMC states are re-sampled in the chain. The higher the number of swaps rejected in the MCMC, the larger the sampling gap of the network. For networks with a high rate of re-sampling, and hence a low rate of change of network structure over time, assessing only a narrow window of degree assortativity values might result in an incorrect test outcome. Hence, using the sampling gap of the network as the window size of the DFGLS test ensures that the number of states used to assess whether convergence has been reached is specific to both the structure of the network and the sampling behaviour of the MCMC in the relevant graph space. Once convergence is detected by the DFGLS test, \hbox{every} $\eta_0$ steps beyond the point of convergence provides effectively an iid draw from the configuration model.


To illustrate this procedure, Fig.~\ref{ConvergenceDetection_green} shows the point at which convergence was detected by our test (red line) for a simple network in the stub-labeled space with $n = 2\,642$ nodes and $m = 3\,303$ edges, and for a loopy multigraph in the vertex-labeled space with $n = 2\,837$ nodes and $m = 11,407$ edges.


For the two networks in the above example, the method described here takes about 30 seconds and about three minutes, respectively, to generate 1000 samples from the configuration model on a modern laptop.

\subsubsection{\bf Validating our method}

\begin{figure}[t!]
\centering
    \includegraphics[width=0.48\textwidth]{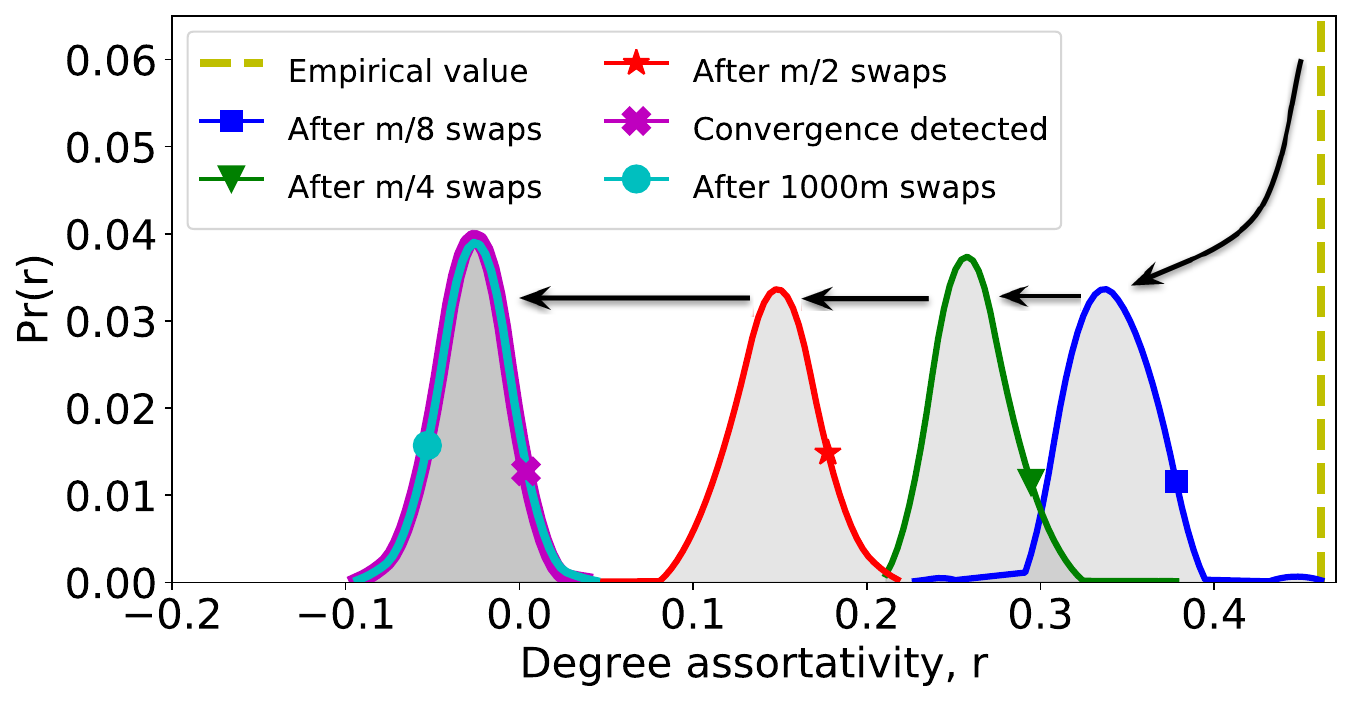}
  \caption{Null distributions obtained after increasing numbers of double edge swaps are applied to a simple network with $n = 1589$ nodes and $m = 2742$ edges in the vertex-labeled space, validating the correctness of our convergence detection method. We find similar results for the other graph spaces.}
\label{Validation}
\end{figure}

If this convergence detection method works as desired, the distribution of assortativity values after convergence is detected should be stationary, meaning that running the Markov chain longer should not alter the sampled assortativity distribution (Fig.~\ref{Validation}). To assess this behavior, we tabulate the assortativity distributions of $500$ states sampled after $m/8$, after $m/4$, and after $m/2$ steps, and then again at convergence and after $1000m$ steps, a point sufficiently deep in the chain that we assume it represents the converged distribution. The $1000m$ distribution provides a comparison against distributions sampled earlier in the Markov chain. 

As a first test, we apply this assessment to a simple vertex-labeled network with $n=1589$ nodes and $m=2742$ edges. Fig.~\ref{Validation} shows the five sampled distributions, along with the assortativity coefficient for the original empirical network, which is the initial condition of the Markov chain. In the pre-convergence phase, as the MCMC walk lengthens from $m/8$, to $m/4$, and then to $m/2$ steps, the assortativity distribution for the sampled networks moves progressively further away from the empirical value of the initial network. This non-stationary behaviour indicates a steady decorrelation of the Markov chain relative to its initial state, as double-edge swaps progressively randomize the network’s structure. Once convergence is detected, this non-stationary behavior is no longer present. Instead the difference between the assortativity distribution at convergence, and well beyond it, is negligible, suggesting the test of convergence correctly detected the Markov chain’s convergence on the target uniform distribution. 

To formally quantify the correctness of this convergence test in detecting the convergence of the MCMC on its stationary distribution, we measure its accuracy on all 509 networks across all the eight graph spaces. For each network in a given graph space, we first obtain a distribution of degree assortativity values at the time of convergence detection by detecting convergence on 200 independent runs of the MCMC, taking one sample from each chain at the point of convergence. We also obtain a distribution of 200 degree assortativity values from 200 independent chains after each chain has been run for $1000m$ swaps (well beyond convergence). We then perform a standard two-sample Kolmogorov-Smirnov (KS) test, with a significance level of $\alpha=0.05$, between the assortativity distribution at the time of convergence and after $1000m$ swaps. 

If there are $D$ networks in a given graph space, we reject the null hypothesis of no early convergence detection by our method in the given graph space if more than $u$ of the $D$ KS tests are statistically significant. The family-wise Type-I error rate for $D$ KS tests is given by \begin{equation}
    \alpha_f = 1 - F_{D, \alpha}(u)\enspace ,
\label{eq:appendix_familywiseTypeIrate1}
\end{equation}
where $F_{D, \alpha}$ is the cumulative distribution function of the binomial distribution \hbox{Bi$(D, \alpha)$}. Since $D =$ 103, 154, 110, and 142, for the simple, loopy, loopy multigraph and multigraph spaces, respectively, we choose $u = 8, 12, 9, 11$ such that the family-wise Type-I error for $D$ KS tests is about 5\%; this choice yields rates of 7.3\%, 4.6\%, 4.9\%, and 5.3\%, respectively. 

\begin{figure}[t]
\centering
    \includegraphics[width=0.48\textwidth]{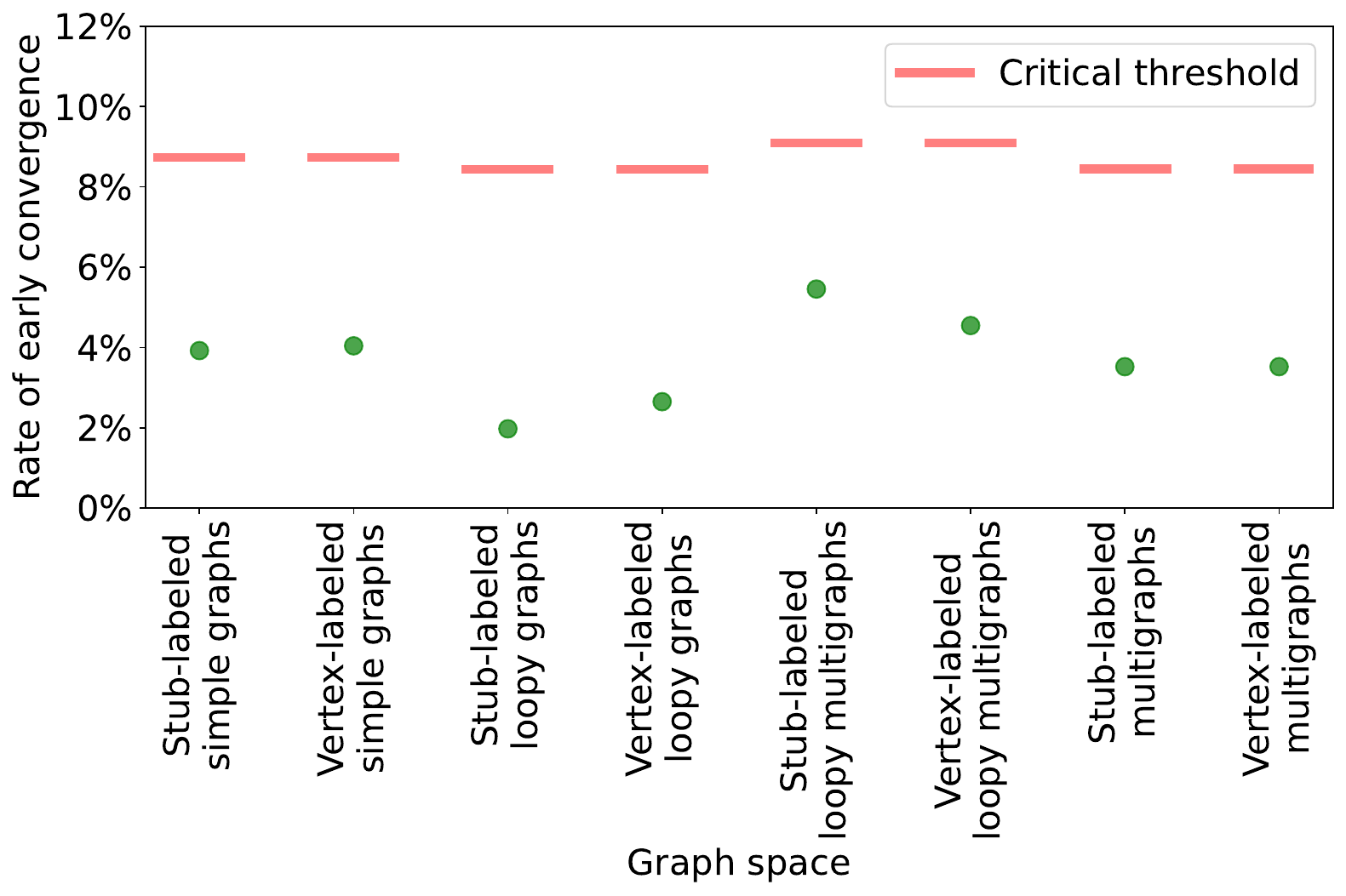}
  \caption{Proportion of networks in each graph space for which a KS test comparing the distribution of degree assortativity values when convergence is detected and that after $1000m$ swaps is statistically significant. The critical threshold of this proportion is chosen such that the family-wise Type-I error is 7.3\%, 4.6\%, 4.9\%, and 5.3\% in the simple, loopy, loopy multigraph, and multigraph space, respectively.}
\label{DFGLS_only_RejectionRate}
\end{figure}

Fig.~\ref{DFGLS_only_RejectionRate} shows the proportion of networks in each graph space for which the null-hypothesis of the KS test is rejected, as well as the critical threshold computed as $(u+1)/D$ for each graph space. Any rejection rate below the critical threshold indicates that our method did not detect convergence too early. Because we obtain rejection rates below the critical threshold in each of the eight graph spaces (Fig.~\ref{DFGLS_only_RejectionRate}), we conclude that our convergence detection method accurately detects when the MCMC has reached its stationary distribution in each of the eight graph spaces.

\subsubsection{\bf Comparing convergence tests}
Detecting convergence in a Markov chain Monte Carlo algorithm is a common and non-trivial statistical problem, especially for scalar time series. Many techniques exist. Although the underlying states in the double-edge swap Markov chain are networks, our approach for detecting convergence converts this graph sequence into a sequence of scalar values. Here, we compare the method described above with three commonly used scalar time series convergence tests: (i) the Geweke diagnostic~\cite{geweke1991evaluating}, (ii) the Gelman-Rubin diagnostic~\cite{gelman1992inference} and (iii) the Raftery-Lewis diagnostic~\cite{raftery1995number} (each of these methods is available via libraries in Python~\cite{pythonpackage} and R~\cite{rpackage}).

A crucial consideration when selecting a convergence test is whether the Markov chain in question satisfies the particular test’s underlying requirements and assumptions~\cite{cowles1996markov}. Each of these three alternative tests make assumptions that may not hold for the configuration model. For instance, the Geweke diagnostic assumes that the Geweke statistic derived from the sampled states will be distributed as a standard normal variable in the asymptotic limit (see Appendix~\ref{appendix:Geweke}). The Gelman-Rubin diagnostic assumes that the MCMC’s stationary distribution of the scalar is normally distributed, and it requires more than one Markov chain to be initialized at highly dispersed initial states in the sample space (see Appendix \ref{appendix:GelmanRubin}). And finally, the Raftery-Lewis diagnostic depends on a quantile; in some cases, the estimated convergence rate may fall far below the rate required of the full Markov chain~\cite{brooks1999miscellanea} (see Appendix \ref{appendix:RafteryLewis}).


\begin{figure}[t!]
\centering
    \includegraphics[width=0.48\textwidth]{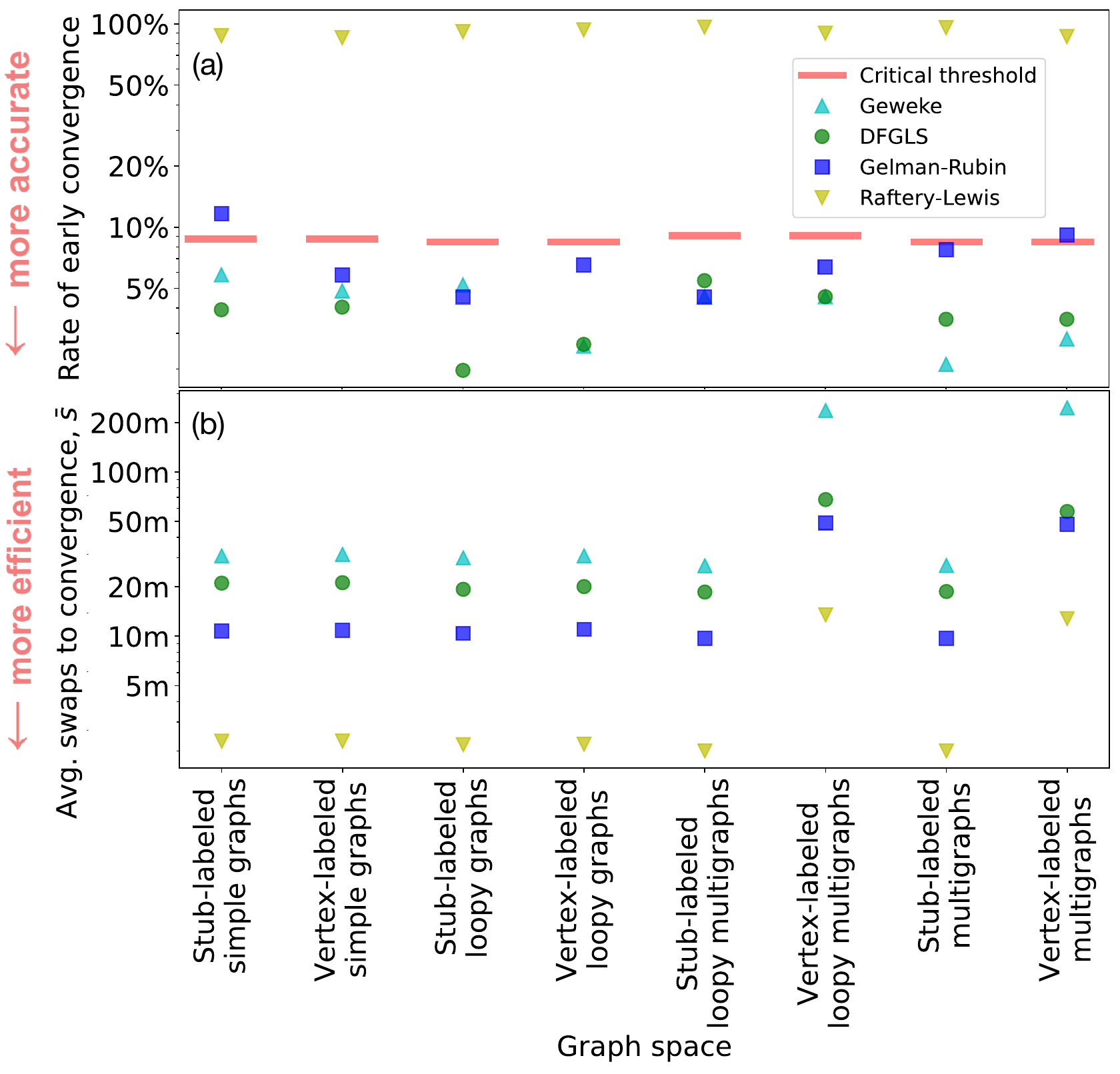}
  \caption{(a) The rate of early convergence (inversely proportional to accuracy), and (b) the average number of swaps applied before convergence is detected (inversely proportional to efficiency) by the four convergence diagnostics applied to all networks in the corpus, across all eight graph spaces.}
\label{ComparisonOtherDiagnostics}
\end{figure}

We evaluate the performance of the four convergence tests according to their accuracy and efficiency. First, we say that a test is accurate if the degree assortativity values at the time that the test detects convergence and the corresponding values at $1000m$ steps into the Markov chain are from the same distribution (i.e., a two-sample KS test is not statistically significant). Second, we say that a test is more efficient than other tests if it detects convergence with fewer double-edge swaps compared to the others. An ideal convergence test will perform well on both accuracy and efficiency measures. An efficient but inaccurate test would allow relatively fewer double-edge swaps, but would tend to declare convergence too early, producing a distribution of assortativity values that differs substantially from the target distribution. In contrast, an inefficient but accurate test would run a very long Markov chain and declare convergence long after the stationary distribution had been reached. In practice, if an ideal test is not available, the latter deviation is preferable to the former.

We apply the four tests to each network in our corpus using a window size equal to the sampling gap of the network. As before, we measure a test’s accuracy by performing a KS test, with $\alpha = 0.05$, and record whether the proportion of networks for which the KS test is statistically significant is within the critical threshold or not. To measure a test’s efficiency, we record the number of steps in the Markov chain (averaged over 200 independent MCMC runs) before each test detects convergence.

Fig.~\ref{ComparisonOtherDiagnostics} shows the rate of early convergence (inversely proportional to accuracy) and the average double-edge swaps $\bar{s}$ applied before convergence is detected (inversely proportional to efficiency) across all networks in the corpus, in all eight graph spaces. In terms of accuracy (Fig.~\ref{ComparisonOtherDiagnostics}a), the DFGLS test performs well along with the Geweke diagnostic: both tests have rates of early convergence less than the critical bounds across all spaces. In contrast, the Gelman-Rubin diagnostic exhibits a rate of early convergence greater than or equal to the critical threshold in two of the eight graph spaces (stub-labeled simple and vertex-labeled multigraph spaces), indicating that in these spaces, it detects convergence before the MCMC has entered its stationary distribution. The Raftery-Lewis diagnostic performs poorly in general ($>90\%$ rate of early convergence). 

\begin{figure*}[t]
\centering
    \includegraphics[width=0.99\textwidth]{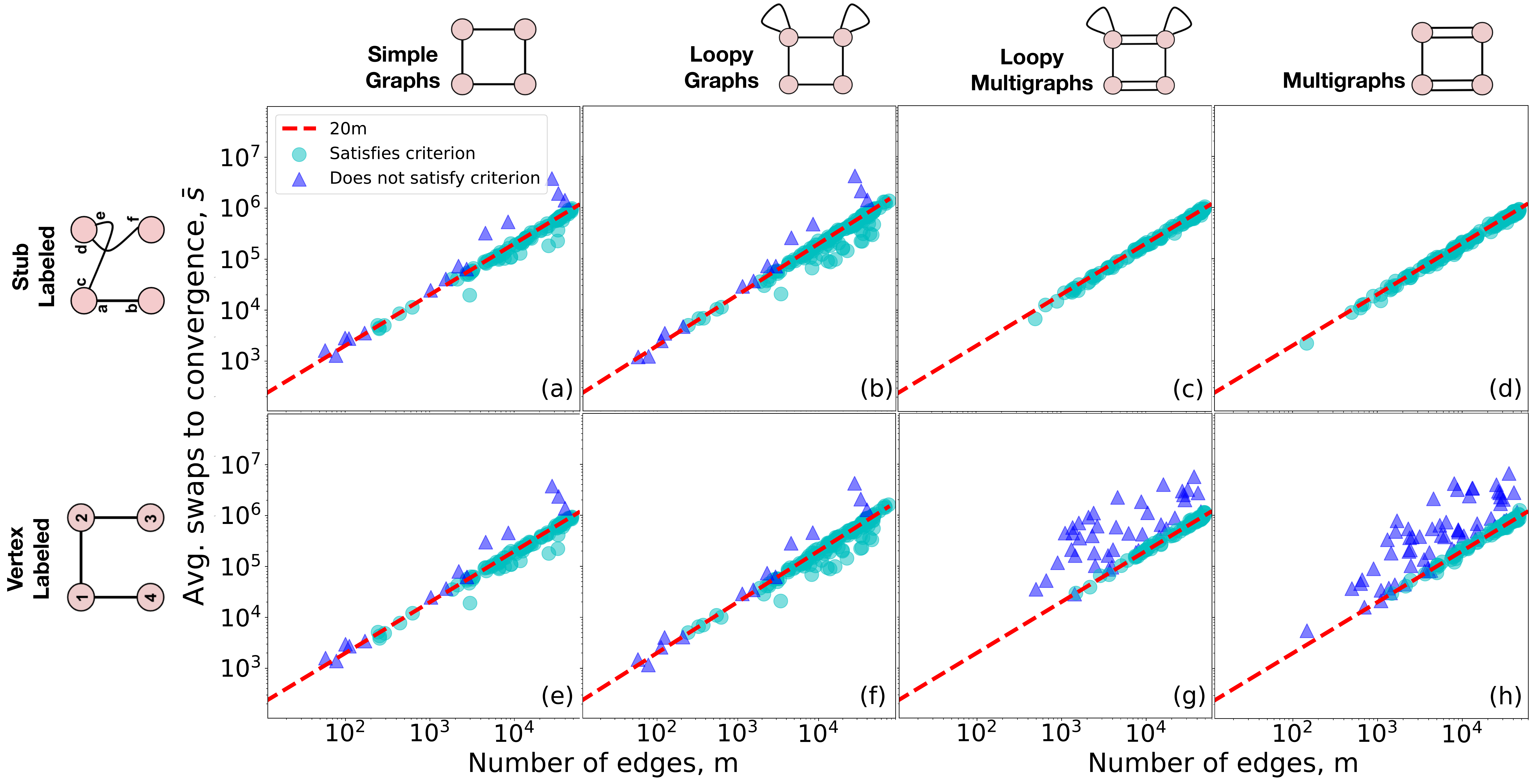}
  \caption{Empirical scaling laws for mixing times of the Fosdick et al. MCMC (averaged over 200 chains). The networks that do not satisfy the maximum density criterion in the simple and loopy graph spaces, or the maximum degree criterion in the vertex-labeled multigraph and loopy multigraph spaces are shown separately (see legend). The red dashed line is a hand-fit line showing an approximation of the central tendency of the mixing times of the networks in all graph spaces except the vertex-labeled loopy multigraph and multigraph spaces (g-h), in which case it captures the central tendency for the networks that satisfy the maximum degree criterion, and serves as a lower bound for the ones that do not.}
\label{AvgSwapsToConvergence}
\end{figure*}

In terms of efficiency (Fig.~\ref{ComparisonOtherDiagnostics}b), the Geweke test is substantially less efficient than the other tests. In fact, in the vertex-labeled multigraph and loopy multigraph spaces, the Geweke test uses on average more than twice as many swaps than the DFGLS test to detect convergence. 

Across both experiments, the DFGLS test exhibits high accuracy and high efficiency, while other techniques tend to perform either slightly or dramatically worse on one or both dimensions. Of the three, the Geweke diagnostic performs most similarly, exhibiting equivalent accuracy, but with far less efficiency (Fig.~\ref{ComparisonOtherDiagnostics}b). The DFGLS test, as described above, is an efficient and practical solution to the problem of detecting convergence, yet is is possible other methods may perform well if optimal parameters for the respective algorithm are identified. This is an interesting direction for future work.


\subsubsection{\bf Mixing time across graph spaces}

To estimate the mixing time of the Fosdick et al. MCMC~\cite{fosdick2018configuring}, we record the number of steps the MCMC runs on average before convergence is detected by our method for each network in our corpus, across the eight graph spaces (Fig.~\ref{AvgSwapsToConvergence}). 

We note that the convergence times of the networks that fail to satisfy the density criterion, i.e., $\rho < 0.134$, in the simple and the loopy graph spaces are much larger than the ones that meet the criterion \hbox{(Fig.~\ref{AvgSwapsToConvergence} a, b, e, f)}. Similarly, in the vertex-labeled multigraph and loopy multigraph  spaces, the convergence times of the networks that fail to satisfy the maximum degree criterion, i.e., $\max_i k_i^2 \leq 2m/3$, are also much larger than the ones that meet it (Fig.~\ref{AvgSwapsToConvergence} g, h). This pattern is consistent with the pattern we observed for sampling gaps, where networks that did not satisfy the density criterion in the simple and loopy graph spaces or the maximum degree criterion in the vertex-labeled multigraph and loopy multigraph spaces have higher sampling gaps than those that did. This similarity in behavior corroborates our suggestion that particular aspects of the network's structure determines the likelihood of re-sampling in the Markov chain, which drives the MCMC's behavior.

In Fig.~\ref{AvgSwapsToConvergence}, we overlay a straight line $\bar{s} = 20m$ in all eight graph spaces to provide a simple low dimensional approximation of the central tendency of the mixing times of the networks across both the stub- and vertex-labeled simple and loopy graph spaces and the stub-labeled loopy multigraph and multigraph spaces. For the remaining two spaces, i.e., the vertex-labeled loopy multigraph and multigraphs, the $\bar{s} = 20m$ line offers an approximation of mixing times for the networks that satisfy the maximum degree criterion, and a lower bound for those that do not. The linear trend between the average swaps to convergence and the size of the networks in our results lead us to conjecture that the convergence time of the Fosdick et al. MCMC~\cite{fosdick2018configuring} is $\Theta(m)$ except in pathological cases. These results may represent a useful insight for theoreticians interested in mathematical bounds on the mixing time of the Fosdick et al. MCMC.

\subsubsection{\bf Extending convergence to other network statistics}

We now consider whether detecting convergence using the degree assortativity implies convergence for other network statistics. For a particular network statistic, we compare its distribution at the point of convergence according to degree assortativity, and a distribution of the same statistic after $1000m$ swaps have been applied. As before, we repeat this for each network across all eight graph spaces. The additional network statistics we explore are:
\begin{itemize}
    \item {\bf Clustering coefficient:} The clustering coefficient of a network is the tendency of the nodes in the network to cluster together in triangles. For simple and loopy graphs, it is defined as the ratio between the number of closed triples and the number of connected triples (both open and closed).
    Since there is no standard definition of clustering coefficient for graphs with multi-edges, we convert the parallel edges of the multigraphs and loopy multigraphs to integer-weighted simple edges where edge weights are the sum of multiplicities of the edges between every pair of nodes, and then we use the definition of clustering coefficient for weighted networks. One can define the clustering coefficient of weighted networks in several ways~\cite{barrat2004architecture,zhang2005general, onnela2005intensity, holme2007korean, saramaki2007generalizations}. Which definition works best for a given scenario depends on the research question being explored. The definition we employ here defines the ``intensity of a triangle" as the normalised geometric mean of the weights of the edges involved in each triangle, and defines the weighted clustering coefficient of each node $i$ of the network as 
    \begin{align}
    \widetilde{C_i} = \displaystyle\frac{2}{k_i(k_i - 1)}\displaystyle\sum_{j, k\neq i}(\widetilde{w}_{ij}\widetilde{w}_{jk}\widetilde{w}_{ki})^{1/3}\enspace,
    \label{weighted_cc}
    \end{align}
    where $k_i$ is the number of neighbours of node $i$ in the weighted version of the network and the edge weights are scaled by the largest weight in the network, $\widetilde{w}_{ij} = w_{ij}/\text{max}(w_{ij})$~\cite{onnela2005intensity}. The weighted clustering coefficient is then averaged over all nodes in the network to obtain the global weighted clustering coefficient. This definition of weighted clustering coefficient ranges between 0 and 1, i.e., $\widetilde{C_i} \in [0, 1]$, which ensures that $\widetilde{C_i}$ equates to the unweighted clustering coefficient when weights are binary, and is invariant to permutation of the weights within a single triangle~\cite{saramaki2007generalizations}.
    
    \item {\bf Diameter}: The diameter of a network is the length of the longest of the geodesic path that exists between any pair of nodes.\ This definition of the diameter produces the same answer regardless of the graph space.
    
    \item {\bf Average path length}: The average path length of a network is the mean of all non-infinite geodesic path lengths among all pairs of nodes in the network. As in case of the diameter, average path length is defined similarly for networks with and without self-loops and multi-edges.
    
    \item {\bf Number of triangles}: We count the number of triangles in a simple and loopy network by counting the number of occurrences of cliques of three distinct nodes. For networks with multi-edges, we convert the network to its weighted version as with the clustering coefficient, and sum the triangle intensities of each triangle, defined similar to Eq.~\eqref{weighted_cc} as the geometric mean of the weights of the triangle, where each weight is normalised by the maximum weight in the network.
    
    \item {\bf Number of squares}: For simple and loopy networks, we count the number of occurrences of groups of four distinct nodes connected in a closed path. For multigraphs and loopy multigraphs, we take the sum of geometric means of the normalised weights of all the squares in a network.
    
    \item {\bf Edge connectivity}: The edge connectivity of a network is defined as the minimum number of edges that need to be removed from the network so that the network becomes disconnected. In networks with multi-edges, the multiplicity of the edges is taken into account.
    
    \item {\bf Radius}: The radius of a network is the smallest of all the node eccentricities of a network, where the eccentricity of a node $v$ is defined as the maximum length of all the shortest paths between the node $v$ and any other node that is reachable from $v$. The same definition of the radius is used for networks with and without multi-edges.

\end{itemize}

\begin{figure}[t!]
\centering
    \includegraphics[width=0.48\textwidth]{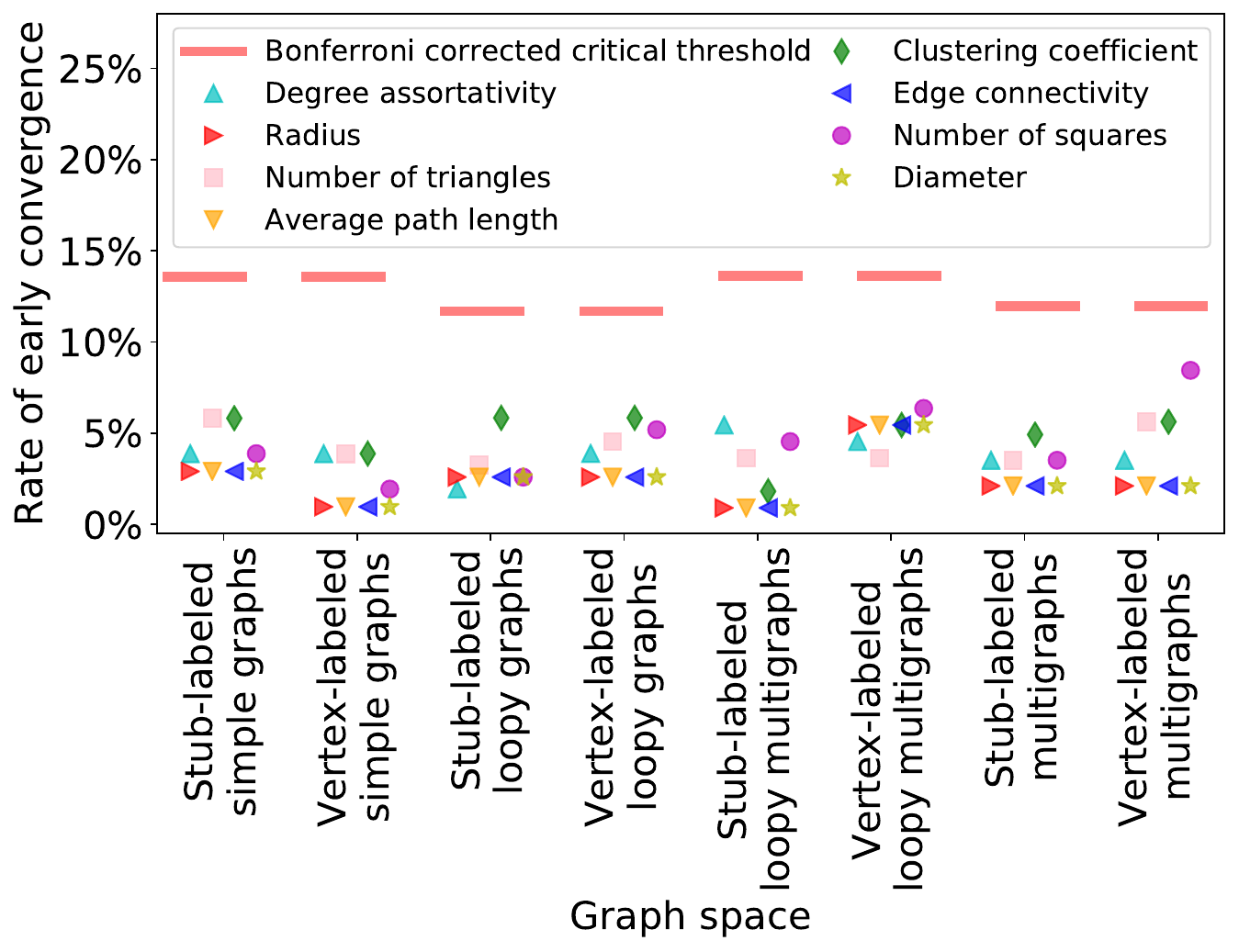}
  \caption{Proportion of networks in each graph space for which the test (either KS-test or the Chi-square test) of whether the distribution of the network statistic (legend) at the point where convergence is detected and that after 1000m swaps have been applied on the network are the same is statistically significant at the 0.05 level of significance.}
\label{ONS_vs_GraphSpace}
\end{figure}

See Appendix~\ref{appendix:Last} for an example on how these network statistics change as swaps are applied. For comparing the distributions at the point of convergence and after $1000m$ swaps, we again use a KS test with $\alpha = 0.05$. If the network statistic has a distribution with less than 10 unique values, we instead apply a chi-square test of independence. Fig.~\ref{ONS_vs_GraphSpace} summarizes the proportion of networks in the corpus for which the tests are rejected across the eight network statistics (including degree assortativity). Because we are performing 64 hypothesis tests to evaluate our method (8 network statistics in each of the 8 graph spaces), we perform the Bonferroni correction to account for multiple testing. To bound our family-wise Type-1 error rate at 0.05, we set the individual test-wise level to 0.05/64 = 0.0007. The corresponding critical thresholds for the rate of early convergence are 13.59\%, 11.69\%, 13.63\%, and 11.97\% in the simple, loopy, loopy multigraph, multigraph spaces, respectively, and are shown in Fig.~\ref{ONS_vs_GraphSpace} as red horizontal bars. Across each of the eight graph spaces, we find that the distributions of all eight of the network statistics have converged on their asymptotic forms when convergence is detected according to degree assortativity, suggesting that degree assortativity converges more slowly than these alternative statistics and hence represents a conservative choice for a convergence test.

\section{Applications}

In this section we use two real-world examples to demonstrate the application of our method on empirical networks. The goal is to evaluate whether an empirically observed network characteristic can be explained as a consequence of the degree sequence alone. 

\subsection{The centrality of the Medici}

\begin{figure}[t!]
\centering
    \includegraphics[width=0.48\textwidth]{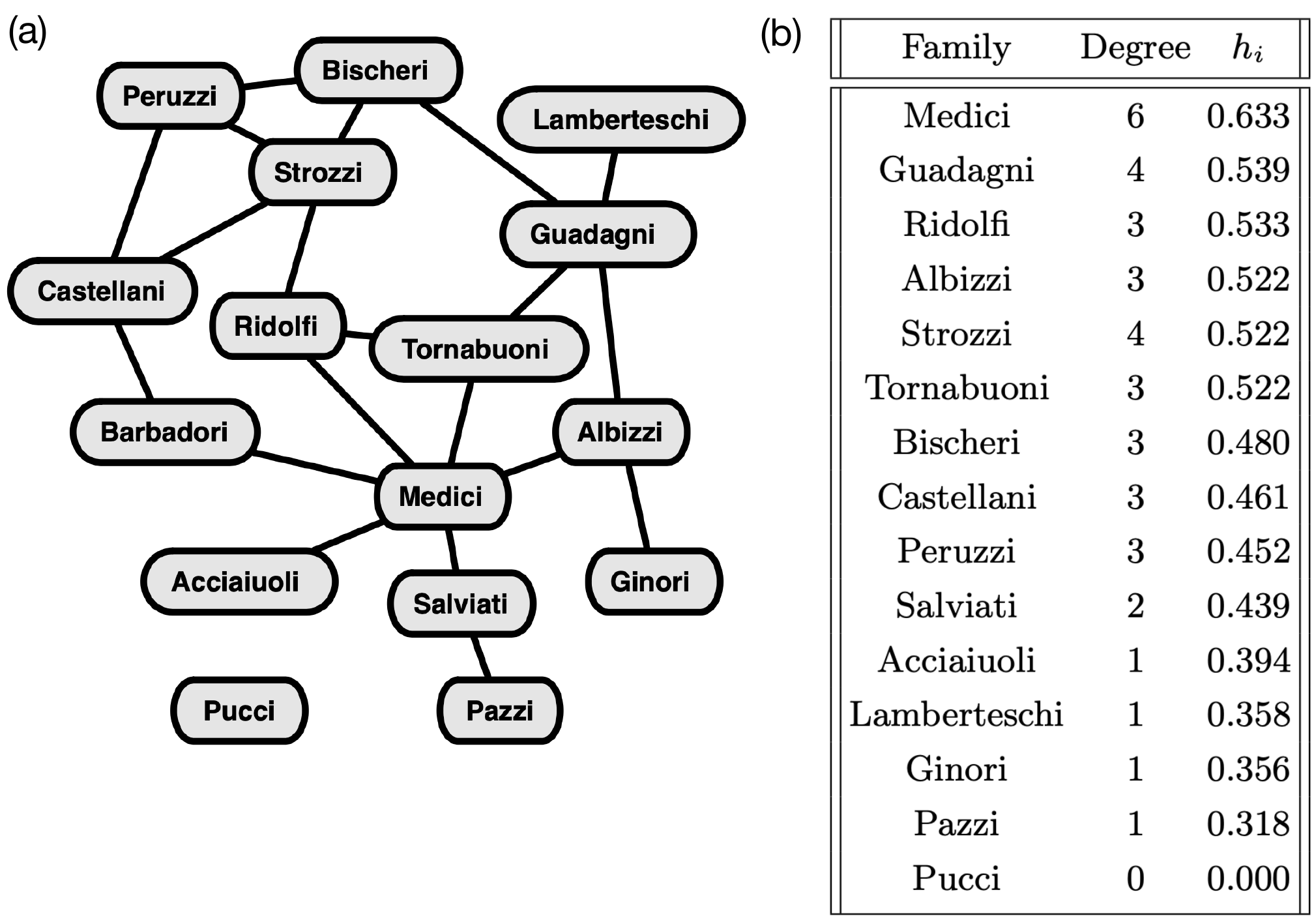}
  \caption{(a) The Medici family marriage network, from Ref.~\cite{padgett1993robust} with $n = 16$ nodes and $m = 20$ edges. (b) Each family's node degree, and harmonic centrality $h_i$ within the Florentine network.}
\label{FlorentineFamilyNetwork}
\end{figure}

In the 15th century, the Medici family rose to become one of the most prominent and powerful families in Florence, Italy. Their support of art and humanism in Florence is believed to have led to the early Renaissance in Europe. The network-based explanation that past studies~\cite{padgett1993robust} have provided for the Medici's rise in power is that they established themselves as the most central family among other elite families in Florence, occupying the most important position structurally, which they leveraged in terms of information flow, business settlements, and political planning~\cite{jackson2010social}. Fig.~\ref{FlorentineFamilyNetwork}a shows the network of marriage ties among 16 key elite Florentine families ($n = 16, m = 20$)~\cite{ICON} whose support or opposition towards the Medicis has been established~\cite{breiger1986cumulated} (note: a broader network of contemporary Florentine families contains 116 nodes~\cite{kent1978rise}). This network is a simple vertex-labeled network where each vertex represents a key Florentine family and two families are connected by an edge if there is any marriage tie between them. It is evident that the Medici had more connections than any other family, including their key competitors the Albizzi and the Strozzi. To quantify how well connected each Florentine family was to the others, we compute the harmonic centrality of each family $i$  (Fig.~\ref{FlorentineFamilyNetwork}b), defined as

\begin{equation}
{h_i} = \displaystyle\frac{1}{n-1}\displaystyle\sum_{j=1;j\neq i}^n \displaystyle\frac{1}{\ell_{ij}}\enspace .
\end{equation}

Under this measure, the Medici family is indeed the most important node in the network ($h_i = 0.633$).
Now, we examine whether the highly central position of the Medici family can be explained by the degree structure of the Florentine network alone. To answer this question, we generate an ensemble of 1000 random networks with the same degree sequence as the Florentine network, and compute the distribution of harmonic centralities for each family in the network. We sample the random networks from the simple vertex-labeled graph space, because in the given setting, marriage ties cannot exist within the same family (i.e., no self-loops), families are connected if their members ever married with each other (i.e., no multi-edges), and stubs do not have distinct labels (i.e., vertex-labeled). 

Given the small size and the structure of the Medici network, one could have generated the reference distribution of simple networks using the repeated stub-matching algorithm as well, i.e., by repeatedly generating networks from the stub-matching algorithm until a network with no self-loops or multi-edges is generated. However, the efficacy of this repeated configuration model for generating uniform draws for simple networks depends in a complicated way on both network size and its degree structure, making it infeasible in a lot of cases. For instance, if the repeated configuration model is run for the Medici network (n = 16, m = 20), the algorithm generates one simple graph in 20 trials on average. However, for the Zachary karate club network (n=34, m=78), the same is generated once in 65 million trials on average. This implies that the repeated configuration model cannot be used as a replacement for our method, even for smaller networks with only a few dozen nodes.

\begin{figure}[t!]
\centering
    \includegraphics[width=0.48\textwidth]{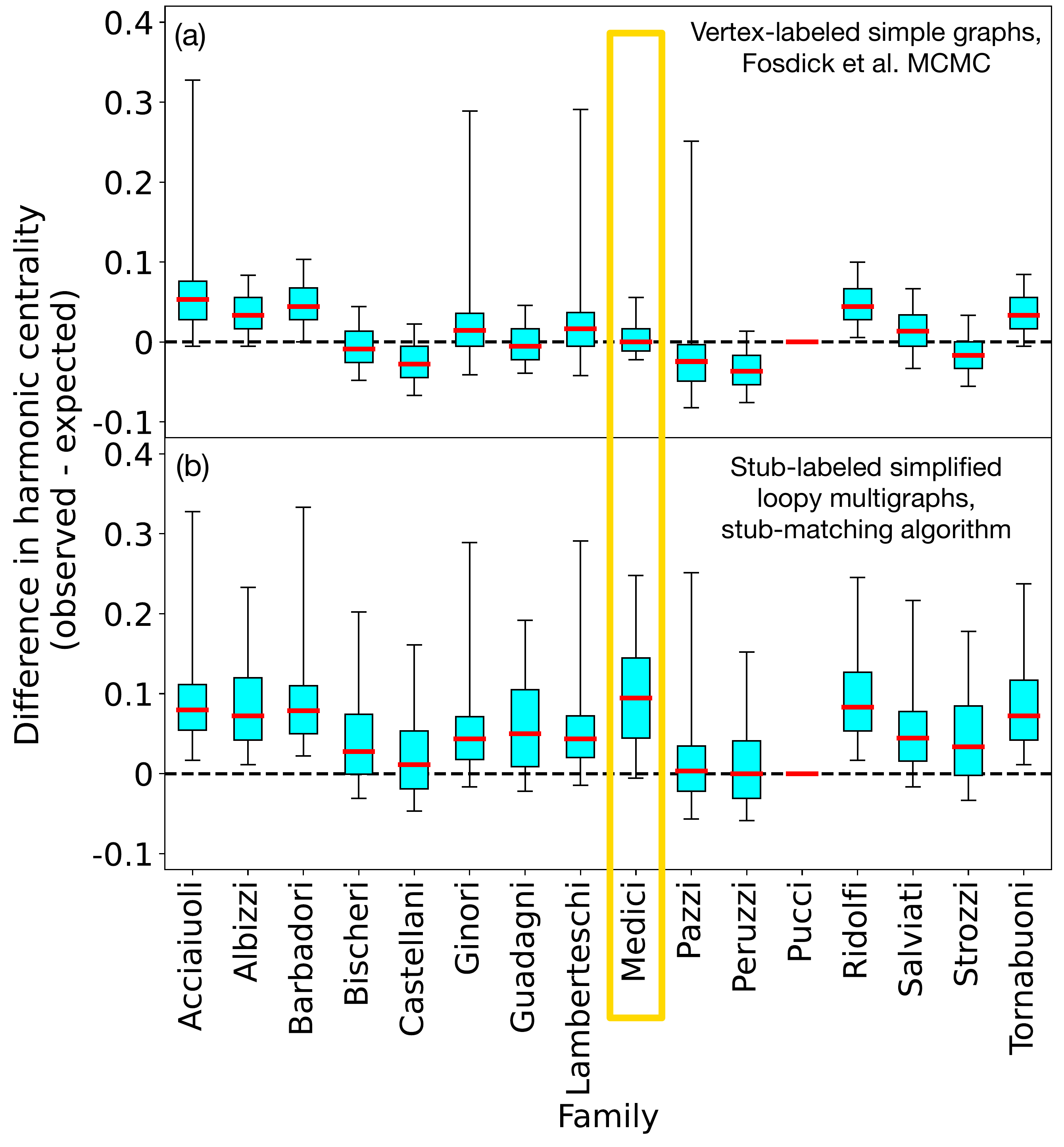}
  \caption{Boxplots illustrating the distribution of difference between the harmonic centrality of each family in the observed Florentine network and its harmonic centrality in the reference ensemble generated (a) from the simple vertex-labeled graph space using our method, and (b) from the stub-labeled loopy multigraph space using the random stub-matching algorithm, where the graphs are simplified by removing self-loops and collapsing multi-edges. The whiskers show the 5th and 95th percentiles and the boxes show the 25th and 75th percentiles with the median value indicated by solid lines; outliers are not shown.}
\label{MediciApplication}
\end{figure}

Fig.~\ref{MediciApplication}a shows the difference between the observed harmonic centrality of each family in the Florentine network and the random values from the corresponding null model ensemble. Notably, the difference between the observed and expected harmonic centrality of the Medici family in the observed network is negligible, indicating that in the network of the elite Florentine families, the centrality of the Medici can be fully explained by the number of marriage ties each Florentine family had in the network.

To illustrate how a conclusion about the explanatory power of the network's degree sequence alone depends on the correct choice of reference distribution, we repeat the analysis, but now generate the reference networks using the commonly used random stub-matching algorithm, followed by collapsing multi-edges and removing self-loops.\ This process of ``simplifying" the sampled loopy multigraphs ultimately changes the null model's degree sequence by introducing a bias, particularly among high degree nodes, as the probability of participating in a self-loop or multi-edge increases with degree.\ Additionally, the random stub-matching algorithm samples networks from the stub-labeled loopy multigraph space, which is an incorrect graph space to sample the reference distribution from in this context. Fig.~\ref{MediciApplication}b shows the difference between the observed and expected harmonic centralities of each Florentine family, using this alternative approach, showing substantially different results compared to using the correct graph space in Fig.~\ref{MediciApplication}a.\ In fact, the results of Fig.~\ref{MediciApplication}b would lead one to incorrectly conclude that the harmonic centrality of the Medici in the observed network cannot be explained by the network's degree structure alone, while Fig.~\ref{MediciApplication}a shows clearly that it can be.\ This exercise illustrates the importance of using the correct method for generating a reference distribution for such evaluations.\ Using the wrong method to generate the reference distribution can lead to conflicting or erroneous conclusions without providing any indication \hbox{to the researcher of the error}.

\subsection{Gender assortativity in Dutch high school emotional support network}


Attribute assortativity or homophily in a network is the tendency of nodes to be connected to others with similar attributes. This form of assortative mixing can be quantified in much the same way that we measure degree assortativity in Eq.~\ref{eq:1}, except that we use a node attribute $a_x$ as the variable of interest rather than the degree:
\begin{equation}
    b = \displaystyle\frac{\sum_{xy} (A_{xy} - k_xk_y/2m)a_xa_y}{\sum_{xy} (k_x\delta(x, y) - k_xk_y/2m)a_xa_y}\enspace ,
\label{eq:7}
\end{equation}
where $a_x$ denotes the scalar node attribute for node $x$. 

\begin{figure}[t!]
\centering
    \includegraphics[width=0.45\textwidth]{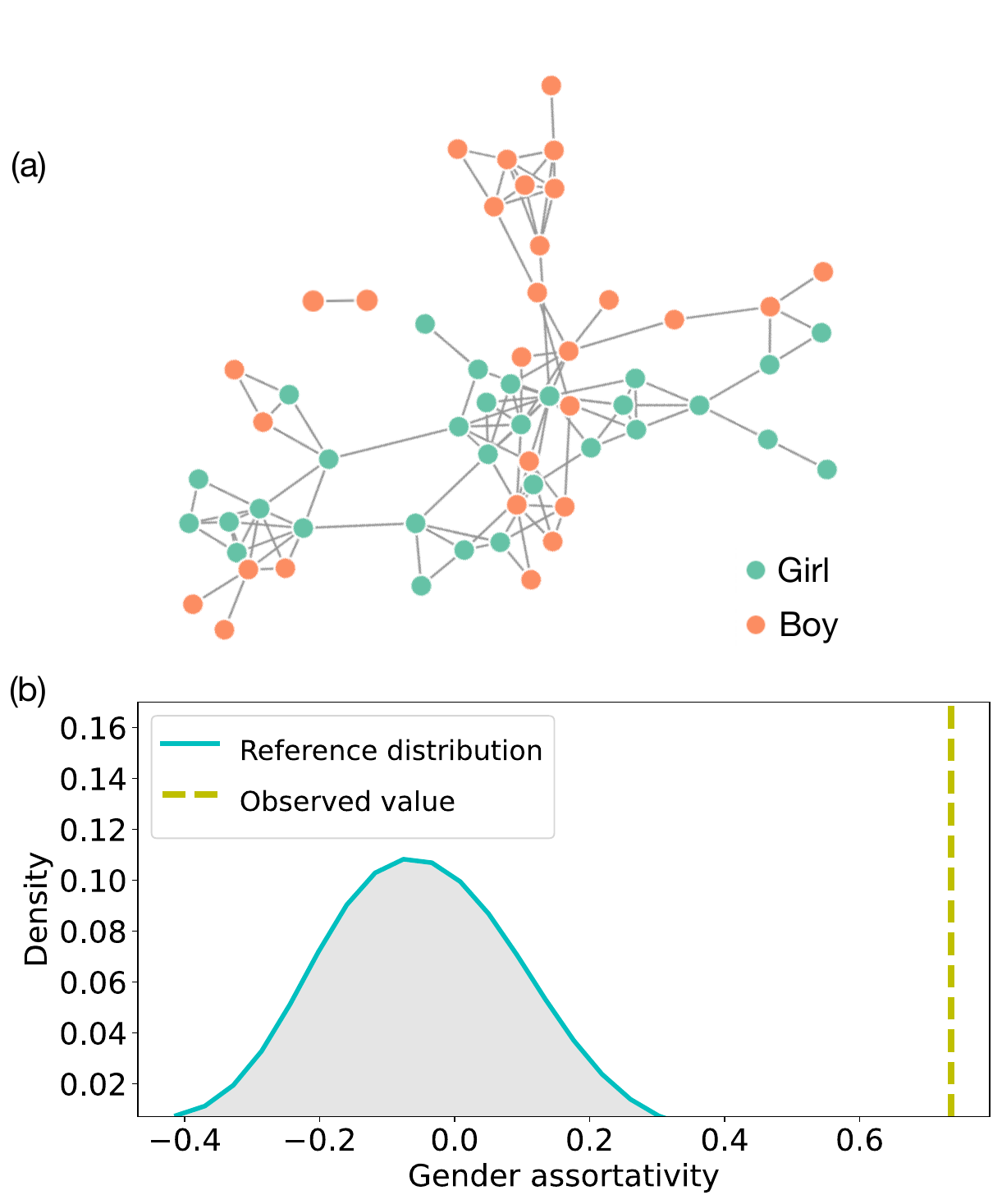}
  \caption{(a) The emotional support network ($n = 73$, \hbox{$m = 51$})~\cite{Wapman2019webweb} of students of a Dutch high school (school no. 23). The gender assortativity coefficient of this network is 0.74. (b) The grey-shaded curve depicts the distribution of gender assortativity coefficient of the networks generated from the simple vertex-labeled graph space of the configuration model, while the dashed-line of the right shows the gender assortativity of the observed network ($p < 0.0001)$. We find evidence that the sharing of emotional support among the students in this Dutch high school is significantly positively correlated with gender.}
\label{DutchSchoolApplication}
\end{figure}

In this example, we consider a survey of fourth grade students in a Dutch urban high school (school no. 23) that participated in the Dutch Social Behavior Study 1994-1996 \cite{houtzager1999just}. In this network (Fig.~\ref{DutchSchoolApplication}a), nodes are fourth grade students and two students are connected by an edge if both students have indicated that they give and/or receive emotional support from each other ($n = 73, m = 51$). Students in this network appear to exhibit a strongly gendered preference when giving and receiving emotional support, with a gender assortativity coefficient $b = 0.74$. 

To assess whether this gender assortativity is merely a consequence of the degree sequence of the network, we generate a reference distribution of $b$ using 1000 networks with the same degrees. We select the graph space to be simple vertex-labeled graphs, because in the study, a student cannot receive support from themselves (i.e., no self-loops), students are connected if they ever took/received emotional support (i.e., no multi-edges), and stubs are indistinguishable (i.e., vertex-labeled). Fig.~\ref{DutchSchoolApplication}b shows the resulting null distribution of gender assortativity coefficients for these reference networks, which indicates that the observed gender assortativity is statistically significant (one-sided $p < 10^{-4}$). Thus, we conclude that the sharing of emotional support among the students cannot be explained merely by the underlying degree structure and distribution of node attributes in this social network, indicating the presence of other social mechanisms. Repeating this analysis on the other schools in the Dutch Social Behavior Study yields a consistent pattern of gendered emotional support exchanges.

\section{Discussion}
The configuration model is among the most widely used models of random graphs in network science. The lack of accurate and efficient methods for generating graphs from the configuration model has discouraged researchers from using it as a null model in empirical research, and has encouraged them to rely on a fast random stub-matching algorithm, that is correct only for generating stub-labeled loopy multigraphs. Heuristics for converting random loopy multigraphs into other types, e.g., simple graphs, introduce structural artifacts that can contaminate empirical conclusions~\cite{fosdick2018configuring}. The methods described here provide a solution to sampling from the configuration model using the Fosdick et al.\ MCMC~\cite{fosdick2018configuring} in eight graph spaces, defined by whether the target graph is stub-labeled or vertex-labeled, and whether it allows self-loops or not, and multi-edges or not.

Our approach transforms the sequence of graphs sampled by the Markov chain into a scalar-valued sequence of degree assortativity values (Fig.~\ref{ChangeIn_r}). We develop a novel algorithm for estimating a sampling gap $\eta_0$ (Algorithm \ref{Algorithm:1}) by which to obtain effectively uncorrelated draws from the Markov chain. This algorithm is based on a standard autocorrelation test to determine how far apart two sampled states must be in order to be statistically independent. Applying this gap estimation algorithm to a large corpus of $509$ real-world and semi-synthetic networks, we identified and organized a set of simple decision rules based on the empirical scaling behavior of estimated gaps $\eta_0$ with the number of edges $m$ (Fig.~\ref{DecisionTree}). These rules allow researchers to automatically select an appropriate sampling gap for a given network usually without having to run the sampling gap estimation algorithm itself. We use the Dickey-Fuller Generalised Least Squares (DFGLS) test to assess stationarity in this Markov chain and show that the test accurately and efficiently detects convergence in all eight graph spaces (Fig.~\ref{DFGLS_only_RejectionRate}). The dependence of both the mixing time and the sampling gap on the re-sampling rate of states in the Markov chain suggests that the sampling gap offers a one-parameter summary of the underlying geometry of the space of random graphs that the Markov chain samples from for a particular network. Our results support a conjecture that the mixing time of the Markov chain in all eight graph spaces is $\Theta(m)$ (Fig.~\ref{AvgSwapsToConvergence}). 

Even though degree assortativity $r$ is more computationally efficient to calculate on a potentially long sequence of networks than many alternative network statistics, networks with very low variance in degrees do exist (e.g., a $k$-regular network), and for these the degree assortativity is either undefined, or changes negligibly as the Markov chain progresses. In these cases, a different network statistic should be used to summarize the sequence of sampled graphs, e.g., the clustering coefficient, albeit at a greater computational cost. The methods developed here are only applicable to the classic configuration model on dyadic networks, i.e., networks where edges are defined as pairs of nodes.  As such, different methods may be needed for correctly sampling from the hypergraph configuration model~\cite{chodrow2020configuration}, in which edges are polyadic, the configuration models for simplicial complexes~\cite{young2017construction}, or networks where edge weights cannot be interpreted as multi-edges. Finally, the Fosdick et al.\ MCMC cannot be applied to the loopy networks that do not satisfy the necessary conditions for the loopy graph space to be connected~\cite{nishimura2018connectivity}. Such loopy networks are extremely rare.

Our analysis of the estimated sampling gaps reveals several interesting patterns, along with useful insights on space-specific conditions under which the double-edge swap MCMC tends to re-sample states frequently. We leave exploration of different choices for quantifying the degree of autocorrelation between two sampled states, and whether that may yield more efficient gap estimation algorithms, for future work. Exploring whether a more efficient test than ours could be constructed without compromising its accuracy is another direction for future work. The appearance of scaling laws in the estimated sampling gaps, and the consistency of their form across different graph spaces is intriguing. This pattern may reflect a currently unknown but common underlying topology both within and across these different graph spaces. Investigating the origins of this common pattern may yield deeper theoretical insights on guarantees for MCMC convergence for sampling random graphs. Our study may also provide insights on developing convergence detection methods for other Markov chain algorithms for networks~\cite{nishimura2018connectivity}, and other variations of the configuration model, e.g., on graphs of fixed core-value sequence~\cite{van2021random}.

\begin{acknowledgements}
The authors thank Dan Larremore, Alex Fout, Joel Nishimura, Johan Ugander, Brian Zaharatos, Vanja Dukic, and Sam Zhang for helpful conversations, and Kevin Sheppard for helping with the Python implementation of the DFGLS test. They acknowledge the BioFrontiers Computing Core at the University of Colorado Boulder for providing High Performance Computing resources (NIH 1S10OD012300) supported by BioFrontiers IT. The authors thank Joel Nishimura for providing the Python code for checking if the loopy graph space is connected or not. This work was funded in part by Air Force Office of Scientific Research Award FA9550-19-1-0329.
\end{acknowledgements}

\section*{Author contributions}
UD, BF and AC conceived the research and designed the analyses. UD conducted the analyses. UD, BF and AC wrote the manuscript.  Competing interests: None.

\bibliographystyle{apsrev4-1}
\bibliography{bibliography}

\appendix
\section{Updating degree assortativity in O(1) time}
\label{appendix:deriv_delta_r}

Here, we provide the derivation for Eq.~\eqref{eq:2} using Eq.~\eqref{eq:1}.

As per Eq.~\eqref{eq:1}, 
\begin{equation}
    r = \displaystyle\frac{\sum_{xy}A_{xy}k_xk_y - \sum_{xy}(k_xk_y)^2/2m}{\sum_{xy}k_xk_xk_y\delta(x, y) - \sum_{xy}(k_xk_y)^2/2m}\enspace .
\label{eq1_simp}
\end{equation}

Recall that
\begin{gather*} 
S_\ell = \sum_{xy}A_{xy}k_xk_y \\ 
S_1 = \sum_xk_x = 2m \\
S_2 = \sum_xk_x^2, S_3 = \sum_xk_x^3
\end{gather*}

Now multiplying both numerator and denominator of Eq.~\eqref{eq1_simp} with $2m$, we get

\begin{equation}
    r = \displaystyle\frac{S_1S_\ell - \sum_{xy}(k_xk_y)^2}{S_1\sum_{xy}k_xk_xk_y\delta(x, y) - \sum_{xy}(k_xk_y)^2}\enspace .
\label{eq_r}
\end{equation}

\noindent Since $\delta(x, y) = 1$ if $x = y$, and 0 otherwise, it can be derived that 

\begin{gather*} 
\sum_{xy}k_xk_xk_y\delta(x, y) = k_1^3 + k_2^3 + k_3^3 + \dots + k_n^3 = \sum_xk_x^3 = S_3\\
\sum_{xy}(k_xk_y)^2 = \sum_x(k_i^2)^2 = S_2^2
\end{gather*}

\noindent Therefore, 
\begin{equation}
    r = \displaystyle\frac{S_1S_\ell - S_2^2}{S_1S_3- S_2^2}\enspace .
\label{eq1_app}
\end{equation}

\noindent Hence Eq.~\eqref{eq:1} can be re-written in the form of Eq.~\eqref{eq:3} which is a substantially faster way to calculate the degree assortativity of a network. Now we show the derivation for Eq.~\eqref{eq:2}, which is the fast O(1) update formula.\\

Note that in Eq.~\eqref{eq:1}, the summation in the numerator is over all possible node pairs $x$ and $y$ ($n^2$ terms). Let us now suppose that a swap $\{(x, y), (w, z)\} \rightarrow \{(x, z), (w, y)\}$ takes place as shown in Fig.~\ref{DoubleEdgeSwaps}. This swap  results into the addition of the edges $(x, z)$ and $(w, y)$ as well as the non-edges $(x, y)$ and $(w, z)$, and the removal of the edges $(x, y)$ and $(w, z)$ as well as the non-edges $(x, z)$ and $(w, y)$ for the calculation of $r$, where a non-edge $(i, j)$ means $A_{ij} = 0, A_{ji} = 0$ and an edge means $A_{ij} = 1, A_{ji} = 1$. Hence, using Eq.~\eqref{eq:1}, the change in the \textbf{numerator} of $r$,
\begin{equation}
    \Delta r' = 2\Bigg[k_xk_z + k_wk_y - k_xk_y - k_wk_z\Bigg]\enspace .
    \nonumber
\label{eq1}
\end{equation}

Here the leading factor of 2 appears because the network is undirected. Our earlier derivation shows that when the denominator of $r$ in Eq.~\eqref{eq:1} is multiplied by $2m$, we obtain \hbox{$(\sum_ik_i\times \sum_ik_i^3) - (\sum_ik_i^2)^2$}. Hence,

\begin{align}
D & = \sum_{xy} (k_x\delta(x, y) - k_xk_y/2m)k_xk_y \nonumber\\
& = \displaystyle\frac{(\sum_ik_i\times \sum_ik_i^3) - (\sum_ik_i^2)^2}{2m}\enspace , \nonumber
\end{align}

\noindent and $\Delta r$ can be calculated using $\Delta r'$ as 
\begin{equation}
    \Delta r = \displaystyle\frac{(k_xk_w + k_yk_z - k_xk_y - k_wk_z)\times 4m}{(\sum_ik_i\times \sum_ik_i^3) - (\sum_ik_i^2)^2}\enspace .
\label{eq_final_delta_r}
\end{equation}

\section{The autocorrelation function}
\label{appendix:A}
The autocorrelation function of a sample measures the degree to which its values are serially correlated, so that the greater the serial correlation, the  larger the value of the autocorrelation function~\cite{box2011time}. Mathematically, the autocorrelation $a_\tau$ at lag $\tau$ quantifies the average correlation between a pair of values $x_{t}$ and $x_{t+\tau}$ in the sequence, separated by a lag of $\tau$, and ranges from \hbox{$-1 \leq a_\tau \leq 1$}.  It is defined as
$a_\tau = \displaystyle\frac{R_\tau}{R_0}$,
where 
\begin{gather*}
R_\tau = \displaystyle\frac{1}{T}\displaystyle\sum_{t=1}^{T-\tau}{(x_t -\overbar{x})}{(x_{t+\tau} - \overbar{x})}\enspace ,\\
R_0 = \displaystyle\frac{\displaystyle\sum_{t=1}^{T}(x_t - \overbar{x})^2}{T}\enspace.
\end{gather*}

\noindent where $T$ is the sample size and $1 \leq \tau \leq T-1$. Hence, the autocorrelation $R_h$ is the covariance of the sample, and itself, at a lag of $h$, normalized by the variance of the sample\\

\textbf{Power-analysis:} Consider a list $X_{\eta}$ comprised of degree assortativity values in which consecutive values are the degree assortativities of graphs sampled $\eta$ swaps apart in the Markov chain. Based on empirical investigation, degree assortativity values of large networks in our empirical corpus are roughly Gaussian. Let the size of $X_{\eta}$ be $T$. If the values in $X_{\eta}$ are iid normally distributed values, the lag-1 autocorrelation of $X_{\eta}$ will have mean and variance given by
\begin{equation}
    \mu(a_1) = -\displaystyle\frac{1}{T}\enspace ,
\label{eq:appendix_tau1mean}
\end{equation}
\begin{equation}
    \sigma^2(a_1) = \displaystyle\frac{T^4 - 4T^3 + 3T^2 + 4T - 4}{(T+1)T^2(T-1)^2}\enspace ,
\label{eq:appendix_tau1variance}
\end{equation}


\noindent using $\tau = 1$ in Eq.~\eqref{eq:4} and~\eqref{eq:5}, respectively~\cite{dufour1985some}.

Hence, to assess if the degree assortativity values in $X_{\eta}$ are effectively independent (iid), we need to test whether the lag-1 autocorrelation of $X_{\eta}$ is different than expected under iid sampling. We use Eq.~\eqref{eq:appendix_tau1mean} and Eq.~\eqref{eq:appendix_tau1variance} for critical values assuming a normal approximation as in Ref.~\cite{dufour1985some}. Since each state in the Fosdick et al.\ MCMC is expected to be positively correlated with the next and the following state (except in pathological cases), we use lag-1 autocorrelation tests that are upper-tailed.

In estimating a network's sampling gap, we test for independence using $C$ different lists each of the form $X_{\eta}$, obtained from $C$ independent Markov chains because a single Markov chain may not be representative of the entire graph space. If each test is performed at a Type-I error rate $\alpha$, the number of tests rejected when the null hypothesis is true would follow a binomial distribution Bi$(C, \alpha)$. Suppose we reject the null hypothesis that $X_{\eta}$ consists of iid values if more than $u$ tests are rejected. Then the family-wise Type-I error of $C$ tests is given by 
\begin{equation}
    \alpha_f = 1 - F_{C, \alpha}(u) \enspace ,
\label{eq:appendix_familywiseTypeIrate2}
\end{equation}
where $F_{C, \alpha}$ is the cumulative distribution function of a random variable following binomial distribution \hbox{Bi$(C, \alpha)$}.

Each test with sample-size $T$ and Type-I error rate $\alpha$ detects a lag-1 autocorrelation $w$ with approximate power $P$ given by,
\begin{equation}
    P = 1 - \Phi\Big(\displaystyle\frac{\mu(a_1) - w}{\sigma(a_1)} - Q_\alpha\Big)\enspace ,
\label{eq:appendix_singletestPower}
\end{equation}
where $\Phi$ is the cumulative distribution function of the standard normal distribution, $Q_\alpha$ is the $\alpha^{th}$ quantile of the standard normal distribution, and $\mu(a_1)$ and $\sigma(a_1)$ are given by Eq.~\eqref{eq:appendix_tau1mean} and Eq.~\eqref{eq:appendix_tau1variance} respectively.  The family-wise power $P_f$ for $C$ normality tests is then given by 
\begin{equation}
    P_f = 1 - F_{C, P}(u)\enspace .
\label{eq:appendix_familywiseTypeIrate3}
\end{equation}

In the sampling gap algorithm \hbox{(Algorithm~\ref{Algorithm:1})}, we choose $C = 10, u = 1, \alpha = 4\%, T = 500$, and $w = 0.1$ so that we get a family-wise Type-I error rate $\alpha_f = 5.8\%$ and family-wise power $P_f = 99.9\%$. 

If a network statistic other than the degree assortativity (for instance, clustering coefficient) is used for the sampling gap algorithm, the normality assumption of the statistic should be revisited, and other tests~\cite{dufour1985some} should be used if normality is not met.

\section{Drawbacks of using Effective Sample Size to sample effectively independent networks}
\label{appendix:C}

One might wonder that instead of solving the sampling gap estimation problem using Algorithm \ref{Algorithm:1}, one could compute the effective sample size (ESS) of the graph samples from the Markov chain. This approach can have two versions. In the first, we run the Markov chain and store only the summary statistic used to track the chain, e.g., the degree assortativity or whether the degree assortativity exceeds an empirical threshold, until a desired ESS is achieved. In this case, the researcher can then estimate the mean summary statistic for random graphs with the specified degree sequence, along with an error bound on the mean estimate based on the achieved ESS. However, such a tool has only very narrow applications as the sampling procedure is based upon tracking a single summary statistic, or possibly a handful of summary statistics, that are identified \emph{a priori}. If another statistic were identified later as being of interest, the entire sampling procedure would have to be re-executed since the original graphs were not stored, and the cost of this re-execution may be significantly greater, depending on the computational complexity of the new statistic of interest, e.g., degree assortativity is efficient to calculate in an online fashion, while most path-based statistics are not. Hence, such an ESS approach may be efficient for a single statistic, but has limited utility and does not solve the general network analysis task we aim to accomplish.

In the second, instead of storing only the summary statistics of the networks the Markov chain traverses, we instead store the entirety of every network traversed, which allows a researcher to calculate any network statistic of interest later, from those graphs saved in memory. However, even for moderate-sized networks, this approach explodes the corresponding storage requirements. For instance, for a simple vertex-labeled network with $n = 2539, m = 10455$ (one of the networks from our corpus), we had to store 794,000 consecutive graph samples from the Markov chain in order to obtain an ESS of just $N=100$, requiring 127 GB of memory. Although this approach was ``fast" in the sense of requiring only $76m$ steps in the Markov chain, it trades saved time for wasted space up front, and wasted time downstream, as performing any calculation on the 794,000 stored graphs, to produce an ESS of $N=100$, will increase the computational cost by a factor of x7940, simply due to reading that data. In contrast, our method trades off this extravagant memory cost for a merely modest increased computational cost (due to ``wasting" the roughly 2m steps between samples), storing only the effectively independent samples of the Markov chain. Hence, to obtain a sample size of $N=100$, our method requires only 0.16 GB of memory space, which is nearly 1000x less space than an ESS approach. Because our method was designed to be computationally efficient, our independent-draw approach pays a very modest computational premium for this large space saving: in the above example, our method runs the Markov chain about 2.6x longer (2,091,000 vs. 794,000 swaps), but this amounts to less than 7 seconds of additional time and does not account for the additional computational cost of computing the ESS at regular intervals of the chain.

Thus, for the particular problem we set out to solve, an approach based on ESS would either be extremely narrow or exorbitantly impractical. In contrast, by producing a set of effectively independent network draws from the configuration model, our approach is maximally general for downstream uses and strikes an efficient and careful balance between time and space costs that makes it genuinely useful to researchers.

\section{Dependence of sampling gap on the MCMC transition probabilities}
\label{appendix:B}

\begin{algorithm}[H]
\begin{algorithmic}[1]
\Input {initial graph $G_0$, graph space (simple, multigraph, or loopy multigraph)}
\Output {sequence of graphs $G_i$}
 \For {$i<$ number of graphs to sample} 
 \State choose two edges at random
 \State randomly choose one of the two possible swaps
 \If {edge swap would leave graph space} 
\State re-sample current graph: $G_{i} \leftarrow G_{i-1}$
\Else
\State swap the chosen edges, producing $G_{i}$
\EndIf
\EndFor
\end{algorithmic}
\caption{stub-labeled MCMC \label{alg_stub} }
\label{alg:stub}
\end{algorithm}

For reference, we have presented the stub-labeled and the vertex-labeled double-edge swap MCMC algorithm in Algorithms \ref{alg:stub} and \ref{alg:vertex} adapted from Fosdick et. al~\cite{fosdick2018configuring}. In this appendix, we provide a discussion of how the MCMC’s transition probabilities in different graph spaces govern the respective sampling gaps we obtain using Algorithm~\ref{Algorithm:1}.

\subsection{Simple graphs}
\label{appendix:simple}

In the stub-labeled space, if a double-edge swap would cause the Markov chain to leave the graph space, e.g., by creating a self-loop or a multi-edge, the swap is rejected and the current graph $G_{t-1}$ is re-sampled by the MCMC (Algorithm~\ref{alg:stub}). Otherwise, the swap is accepted, producing the new graph $G_t$.

In the vertex-labeled space, if a double-edge swap would cause the Markov chain to leave the graph space, the current graph $G_{t-1}$ is also re-sampled (Algorithm~\ref{alg:vertex}). Otherwise, the algorithm checks the number of distinct vertices and self-loops in the chosen double edges. In simple graphs, the number of distinct vertices among $u, v, x$ and $y$ would either be 3 or 4, and the multiplicity $w_{ab}$ of any edge $(a,b)$ would be 1, because multi-edges are forbidden in the simple graph space. Hence, in Algorithm~\ref{alg:vertex}, the variable $B/A$ will be greater than or equal to 1, and the variable $P$ in the algorithm will always be 1. As a result, if a double-edge swap does not cause the Markov chain to leave the vertex-labeled graph space, it will always be accepted, producing the new graph $G_t$. 

Hence, for simple graphs, the transition probabilities are the same for both the stub-labeled and the vertex-labeled space, and thus so are the induced null distributions too. 

\begin{algorithm}[H]
\begin{algorithmic}[1]
\Input {initial graph $G_0$, graph space (simple graph, multigraph, or loopy multigraph)}
\Output{sequence of graphs $G_i$}
\For{$i<$ number of graphs to sample} 
\State choose two distinct edges $(u,v)$ and $(x,y)$ uniformly at random
\If{Unif$(0,1)<0.5$ }
	\State $u,v\leftarrow v,u$ 
\EndIf
\If {edge swap would leave the graph space} 
	\State re-sample current graph: $G_{i} \leftarrow G_{i-1}$
\EndIf
\If {$\exists$ 4 distinct vertices in $u,v,x,y$}
	\State $A\leftarrow w_{uv}w_{xy}$ 
	\State $B \leftarrow (w_{ux}+1)(w_{vy}+1)$ 
\ElsIf{$\exists$ 3 distinct vertices in $u,v,x,y$}
	\If {$u=v$ or $x=y$}
    		\State $A \leftarrow 2w_{uv}w_{xy}$ 
    		\State $B \leftarrow (w_{ux}+1)(w_{vy}+1)$ 
	\Else
    		\State $A \leftarrow w_{uv}w_{xy}$ 
    		\State $B \leftarrow 2(w_{ux}+1)(w_{vy}+1)$ 
	\EndIf
\ElsIf {$\exists$ 2 distinct vertices in $u,v,x,y$}
	\If {only one of $(u,v)$ or $(x,y)$ is a self-loop}
		\State $G_{i} \leftarrow G_{i-1}$ 
		\State continue 
\ElsIf {both $(u,v)$ and $(x,y)$ are self-loops}
		\State $A\leftarrow 2w_{uu}w_{xx}$ 
    		\State $B \leftarrow \frac{1}{2}(w_{ux}+2)(w_{ux}+1)$ 
	\Else
		\State $A\leftarrow \frac{1}{2}w_{uv}(w_{uv}-1)$ 
		\State $B \leftarrow 2(w_{uu}+1)(w_{vv}+1)$ 
	\EndIf
\Else
	\State $G_{i} \leftarrow G_{i-1}$ 
	\State continue 
\EndIf
\State $P\leftarrow \min \Big(1,\displaystyle\frac{B }{A}\Big)$
\If {Unif$(0,1)<P$}
	\State {swap $(u,v),(x,y)\leadsto(u,x),(v,y) $ to produce $G_{i}$}
\Else
	\State $G_{i} \leftarrow G_{i-1}$ 
\EndIf
\EndFor
\end{algorithmic}
\caption{vertex-labeled MCMC}
\label{alg:vertex}
\end{algorithm}

\subsection{Loopy graphs}
\label{appendix:loopy}
In the stub-labeled space, if a double-edge swap introduces a multi-edge in the graph (which is forbidden), the current graph $G_{t-1}$ is re-sampled. Otherwise, the swap is accepted, producing the new graph $G_t$. 

In the vertex-labeled space, if a double-edge swap would cause the Markov chain to leave the graph space, $G_{t-1}$ is re-sampled, otherwise, Algorithm~\ref{alg:vertex} performs the following check. If none of the edges chosen for the swap are self-loops, then \mbox{$P = 1$} and the swap is accepted. But, if either of the chosen edges is a self-loop, then in Algorithm~\ref{alg:vertex}, we set $A = 2$ and $B \in {1, 2, 4}$. When $B = 1$, then $P = 1/2$, and the swap will be rejected half the time. For all other values of B, we have $P = 1$, and hence the swap is accepted. 

In our corpus of 161 multigraphs and loopy multigraphs, 86 (53\%) of these networks have less than 5\% of edges as self-loops, and 129 (80\%) of them have less than 20\% of edges as self-loops. In fact, in any loopy network with $n$ nodes and $m$ edges, there can be a maximum of $n$ edges that are self-loop and a minimum of $m - n$ edges that are not. Since for most networks $m >> n$, a randomly chosen edge in a loopy network is more likely to not be a self-loop than be one. Therefore, if a proposed double-edge swap does not move the Markov chain out of the graph space, $P$ assumes the value 1 with high probability in Algorithm~\ref{alg:vertex}. Hence, the Markov chain behaves very similarly in the stub-labeled and vertex-labeled spaces for loopy graphs.

\subsection{Loopy multigraphs}
\label{appendix:loopymultigraphs}
In the stub-labeled space, every double-edge swap is allowed. Hence, swaps are never rejected and the graph $G_{t-1}$ is never re-sampled.

However, in the vertex-labeled space, whether the proposed double-edge swap is accepted or rejected depends wholly on the multiplicity of the edges and the number of distinct vertices chosen for the double-edge swap. In Algorithm~\ref{alg:vertex}, there can be several ways in which the value of the variable $P$ is far below 1 and hence the graph $G_{t-1}$ will be re-sampled. For this reason, the MCMCs for stub-labeled and vertex-labeled loopy multigraphs  behave differently, with a substantial number of re-samplings occuring in the vertex-labeled space, and none in the stub-labeled space. 

\subsection{Multigraphs}
\label{appendix:multigraphs}
In both the stub-labeled and the vertex-labeled spaces, a graph is re-sampled only if a double-edge swap would introduce a self-loop in the network. 
However, in the vertex-labeled space, re-sampling can occur for an additional reason (Algorithm~\ref{alg:vertex}), depending on the multiplicity of the edges chosen for the double edge swap. In this case, the Markov chain in the stub-labeled and the vertex-labeled spaces have different transition probabilities, which leads to different sampling gaps between the two spaces. 

\section{Probability of rejection due to multi-edges and self-loops}
\label{appendix:Rejection_multi_vs_loopy}

\begin{figure}[t!]
\centering
    \includegraphics[width=0.48\textwidth]{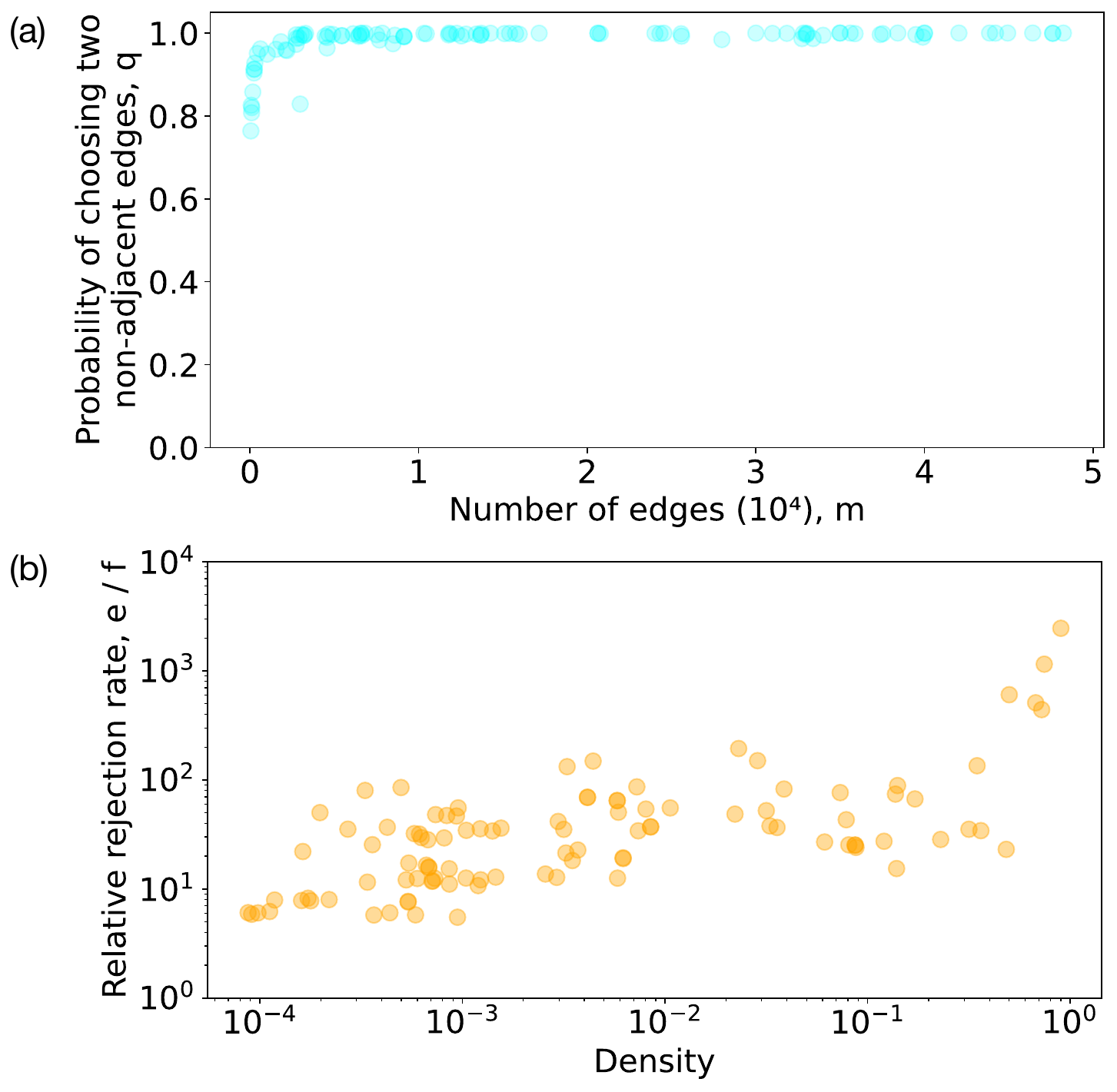}
  \caption{(a) Probability $q$ that two edges chosen uniformly at random for a double-edge swap are non-adjacent, calculated for all 103 simple networks in our corpus using Eq.~\eqref{eq:q}. $q\approx 1$ (b) Ratio $e/f$, where $e$ is the rate at which a swap is rejected due to the introduction of a multi-edge and $f$ is that due to a self-loop. Rejection due to multi-edges is 5-2500 times more than that due to self-loops, and this ratio increases as a network's density grows.}
\label{Rejection_multi_vs_loopy}
\end{figure}

We derived the probability that a double-edge swap is rejected because it introduces a multi-edge in a network that is forbidden by the graph space is approximately $q \times (2\rho - \rho^2)$ (Eq.~\eqref{eq:8}), where $q$ is the probability of two randomly chosen edges being non-adjacent to each other (Eq.~\eqref{eq:q}) and the probability of an edge existing between two randomly chosen edges is approximately the network edge density $\rho$. \hbox{Fig.~\ref{Rejection_multi_vs_loopy}} shows the value of $q$ for all 103 simple networks in our corpus. It is evident that $q\approx 1$ for almost all these networks. Hence, we can approximate the probability that a double-edge swap is rejected due to the introduction of a multi-edge as $(2\rho - \rho^2)$.

Next, we calculate the probability that a double-edge swap is rejected due to the introduction of a self-loop when the graph space forbids them. When two edges chosen for a swap are adjacent to each other (Fig.~\ref{DoubleEdgeSwaps}b), one of the two possible swaps introduces a self-loop (probability = 0.5), while the other one does not. Therefore, \textrm{Pr\big(rejection due to self-loop\big)} = $(1 - q)/2$. Since $q \approx 1$ (Fig.~\ref{Rejection_multi_vs_loopy}a), \textrm{Pr\big(rejection due to self-loop\big)} $\approx 0$. Thus, even if a graph space forbids self-loops, the probability of a swap being rejected due to the introduction of one is almost negligible. 

To test the relative rejection rates due to multi-edges and self-loops, for each of the 103 simple networks in our corpus we let the MCMC run for a burn-in period of $1000m$ swaps and we then let it run for another $1000m$ swaps recording what fraction of those swaps are rejection due to a multi-edge $(e)$ or a self-loop $(f)$ (Fig.~\ref{Rejection_multi_vs_loopy}b). Rejection due to multi-edges is 5-2500 times more than the rate due to self-loops. The increase in the ratio $e/f$ with increasing edge density is expected since the higher the density of the network, the higher the probability that an edge already exists between nodes selected for a double-edge swap. 

\section{Common MCMC convergence tests}
PyMC~\cite{pythonpackage} is a Python package for Bayesian statistical modeling and advanced MCMC algorithms, and it provides implementations of three common convergence tests for generic MCMCs: the Geweke diagnostic~\cite{geweke1991evaluating}, the Gelman-Rubin diagnostic~\cite{gelman1992inference}, and the Raftery-Lewis diagnostic~\cite{raftery1995number}.

\subsection{The Geweke diagnostic}
\label{appendix:Geweke}

The Geweke convergence diagnostic~\cite{geweke1991evaluating} compares the mean and the variance of samples from the beginning and the end of a single chain of a MCMC walk. Geweke~\cite{geweke1991evaluating} takes the beginning section to be the first 10\% of the chain and the ending section to be the last 50\%. The test uses the Geweke statistic, defined as the difference between the means of the two samples divided by the standard error. These statistics are estimated using the spectral densities of the two samples evaluated at zero, which takes into account the autocorrelations in the samples. The method uses the departure of the Geweke statistic from the standard normal assumption as an indicator of convergence failure. However, the test statistic is calculated under the assumption that when the MCMC reaches its stationary distribution, the two chain samples will be distributed according to a standard normal, in the asymptotic limit. The test also assumes that the spectral density of the time series has no discontinuities at frequency zero~\cite{cowles1996markov}. Whether these assumptions are satisfied depends on the process that produces the time series. The Geweke diagnostic is known to be very sensitive to the chosen spectral window~\cite{cowles1996markov}. Hence, the behavior of the Geweke test may depend on the degree to which properties of the stationary distribution of the particular MCMC are known.

\subsection{The Gelman-Rubin diagnostic}
\label{appendix:GelmanRubin}
The Gelman-Rubin diagnostic~\cite{gelman1992inference} compares the within-sequence variability and between-sequence variability of multiple sequences (at least two) obtained from MCMCs with starting points sampled from an overdispersed distribution. This test is based on the idea that when an MCMC has not yet converged, the variance within each chain is much less than that between the chains, because prior to convergence, the MCMC samples states non-uniformly. If $\theta$ denotes the sequence of an MCMC chain, then first, an estimate of the marginal posterior of Var($\theta$) is calculated as a function of the within-sequence variance $(W)$ and between-sequence variance $(B)$ of the multiple sequences. If the MCMC walks have not yet converged, $B$ would overestimate Var($\theta$) because the walks' starting values were chosen to be overdispersed, whereas $W$ would underestimate Var($\theta$) because the MCMC walks have not yet saturated the states in the stationary distribution. 

The Gelman-Rubin statistic is then given by the square root of the ratio of Var($\theta$) and $W$. In the stationary distribution of the MCMC, both Var($\theta$) and $W$ should approach the true variance of the MCMC chain, and hence values of the Gelman-Rubin statistic close to 1 indicate convergence. In general, practitioners often use a cutoff of 1.1 as an indicator of convergence. This convergence test assumes that the stationary distribution of the MCMC is normal. Cowles and Carlin~\cite{cowles1996markov} suggest that both the assumption of a normal approximation of the target distribution, and  the requirement of multiple MCMC chains with highly dispersed initial conditions, may not be reasonable in most practical situations.

\subsection{The Raftery-Lewis diagnostic}
\label{appendix:RafteryLewis}
The Raftery-Lewis diagnostic~\cite{raftery1995number} is based on two-state Markov chain theory and standard sample size formulae for binomial variance. In particular, it calculates the burn-in period of the Markov chain and the total number of subsequent iterations required to accurately estimate $u$, a $q$-quantile of the MCMC’s posterior distribution. The value of $q$, the margin of error $r$, and the probability $s$ of obtaining the estimate in the interval $(q-r,q+r)$, are all user-defined parameters, although the default value of $s$ is 95\%. The method also calculates a thinning interval $k$, which is the number of iterations that should be skipped to produce a chain of independent samples from the Markov chain (analogous to $\eta_0$ in our method).

From the Markov chain $\{\theta_t\}$, the Raftery-Lewis method first constructs a 0-1 binary chain $\{Z_t\}$, and then chooses the thinning interval $k$ to be the smallest natural number for which a first-order Markov chain model of the thinned out chain is statistically preferred over the second-order Markov chain model. For convergence detection purposes, we are interested only in the burn-in period's value, which provides an estimate of the number of steps needed before the MCMC reaches its stationary distribution~\cite{brooks1998convergence}. A detailed analysis of the Raftery-Lewis method by Brooks and Roberts~\cite{brooks1999miscellanea} shows that the method depends strongly on the quantile of interest $q$ and it does not provide information about the chain as a whole~\cite{brooks1999miscellanea}. Furthemore, in certain cases, the convergence rate estimated by this method is far below the convergence rate of the full chain~\cite{brooks1999miscellanea}. As a general rule, they find that the routine value of $q = 0.025$ commonly used in the literature should not be used, as it tends to underestimate the true convergence time. Instead, they suggest that for practical purposes, the diagnostic could be applied to several different $q$ values and then choose the quantile that estimates the largest burning length.

\section{Other commonly used network statistics}
\label{appendix:Last}
\begin{figure*}[t!]
\centering
    \includegraphics[width=0.98\textwidth]{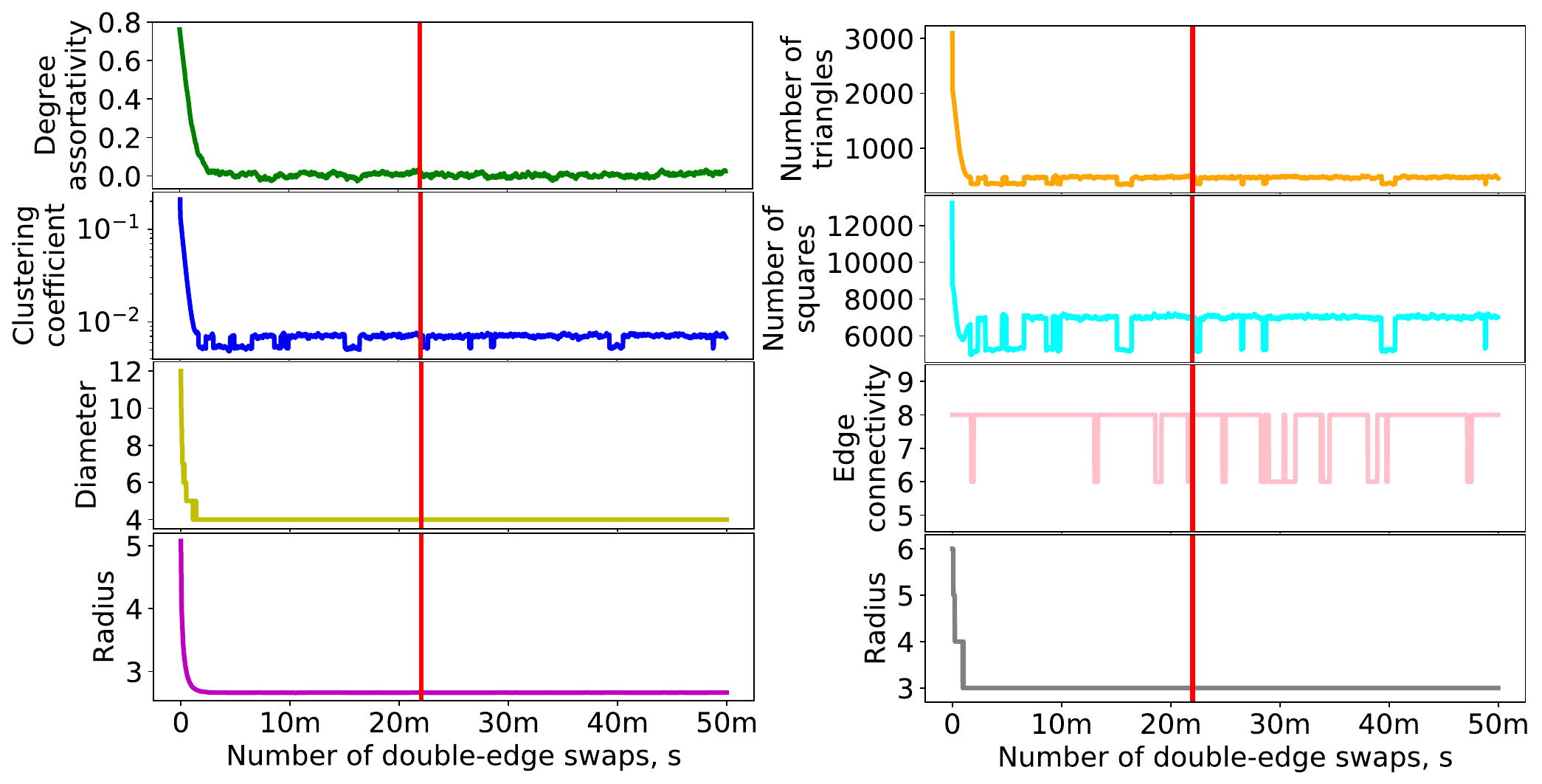}
  \caption{Eight different network statistics as a function of the number of double-edge swaps performed for a vertex-labeled loopy multigraph $n=1015$ nodes and $m = 9988$ edges. The red line denotes the point where our method detects convergence based on degree assortativity. Each statistic is calculated at intervals of 100 double-edge swaps.}
\label{ONS_vs_time}
\end{figure*}
In this section, we provide an example showing how the network statistics beyond the degree assortativity, i.e., the clustering coefficient, diameter, average path length, number of triangles, number of squares, edge connectivity and radius, change as double-edge swaps are applied on a vertex-labeled network with 1015 nodes and 9988 edges (Fig.~\ref{ONS_vs_time}). Since some of the network statistics are computationally expensive to compute and the network size is fairly large, we calculate the statistics once after every 100 double-edge swaps. Each network statistic's value moves away from the initial value as more double-edge swaps are applied. After sufficient swaps are applied, the network statistic converges towards a value, after which it displays non-trivial fluctuations around it. 


\end{document}